\documentclass[lettersize,journal]{IEEEtran}
\usepackage{amsmath,amsfonts}
\usepackage{algorithmic}
\usepackage{algorithm}
\usepackage{array}
\usepackage{multirow}
\usepackage{makecell}
\usepackage[caption=false,font=normalsize,labelfont=sf,textfont=sf]{subfig}
\usepackage{textcomp}
\usepackage{stfloats}
\usepackage{url}
\usepackage{verbatim}
\usepackage{booktabs}
\usepackage{graphicx}
\usepackage{hyperref}
\hypersetup{
    colorlinks=true,
    linkcolor=blue,
    urlcolor=blue,  
    citecolor=blue}

\begin{document}

\title{Frequency-Enhanced Hilbert Scanning Mamba for Short-Term Arctic Sea Ice Concentration Prediction}

\author{
    Feng Gao, \emph{Member, IEEE},
    Zheng Gong, 
    Wenli Liu,
    Yanhai Gan, 
    Zhuoran Zheng, \\
    Junyu Dong, \emph{Member, IEEE},
    Qian Du, \emph{Fellow, IEEE}

\thanks{This work was supported in part by the National Science and Technology Major Project of China under Grant 2022ZD0117201, in part by the Natural Science Foundation of Shandong Province under Grant ZR2024MF020. \textit{(Corresponding author: Yanhai Gan)}

Feng Gao, Zheng Gong, Chuanzheng Gong, Yanhai Gan, and Junyu Dong are with the State Key Laboratory of Physical Oceanography, Ocean University of China, Qingdao 266100, China. 

Zhouran Zheng is with the School of Cyber Science and Technology, Sun Yat-Sen University, Shengzhen 518197, China.

Qian Du is with the Department of Electrical and Computer Engineering, Mississippi State University, Starkville, MS 39762 USA.}}

\markboth{IEEE Transactions on Geoscience and Remote Sensing}{Shell}

\maketitle

\begin{abstract}

While Mamba models offer efficient sequence modeling, vanilla versions struggle with temporal correlations and boundary details in Arctic sea ice concentration (SIC) prediction. To address these limitations, we propose Frequency-enhanced Hilbert scanning Mamba Framework (FH-Mamba) for short-term Arctic SIC prediction. Specifically, we introduce a 3D Hilbert scan mechanism that traverses the 3D spatiotemporal grid along a locality-preserving path, ensuring that adjacent indices in the flattened sequence correspond to neighboring voxels in both spatial and temporal dimensions. Additionally, we incorporate wavelet transform to amplify high-frequency details and we also design a Hybrid Shuffle Attention module to adaptively aggregate sequence and frequency features. Experiments conducted on the OSI-450a1 and AMSR2 datasets demonstrate that our FH-Mamba achieves superior prediction performance compared with state-of-the-art baselines. The results confirm the effectiveness of Hilbert scanning and frequency-aware attention in improving both temporal consistency and edge reconstruction for Arctic SIC forecasting. Our codes are publicly available at \url{https://github.com/oucailab/FH-Mamba}.

\end{abstract}

\begin{IEEEkeywords}
Arctic sea ice, Vision Mamba, Hilbert scanning, Sea ice concentration, Spatio-temporal sequence prediction
\end{IEEEkeywords}

\section{Introduction}

\IEEEPARstart{A}{s} a critical component of the global climate system, Arctic sea ice plays a fundamental role in regulating both regional and global temperatures through the albedo effect. By reflecting incoming solar radiation, it helps maintain the Earth’s energy balance \cite{parkinson2021sea}. In recent decades, Arctic sea ice has been melting at a rate significantly faster than previously anticipated, signaling alarming shifts in climate patterns. Accurate prediction of Arctic sea ice dynamics is therefore essential for advancing our understanding of climate change and ensuring safe navigation in the Arctic. Moreover, the development of reliable forecasting models is becoming increasingly important for geopolitical planning and long-term environmental stewardship in this rapidly changing region \cite{tepes2021changes}.

\begin{figure}[t]
\centering
\includegraphics[width=0.7\linewidth]{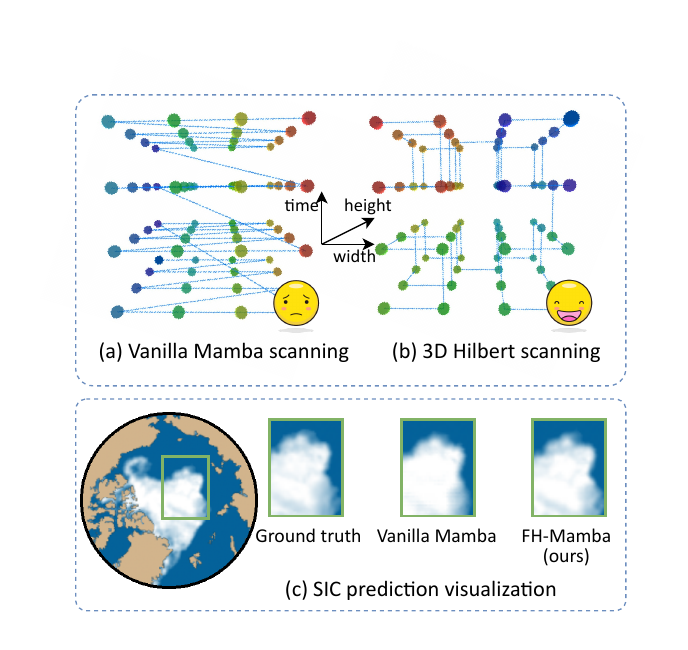}
\caption{Comparison of vanilla Mamba and 3D Hilbert scanning. (a) Scanning route by vanilla Mamba. (b) Scanning route by 3D Hilbert. (c) Visualization of SIC prediction. The lines and endpoints are shaded in gradients from red to green, representing the route of the scan. Vanilla Mamba struggles to effectively model temporal correlation of SIC sequences, while our FH-Mamba leverages the 3D Hilbert curve's locality characteristic to enhance the spatiotemporal learning.}
\label{fig:mot}
\end{figure}

Existing sea ice concentration (SIC) prediction methodologies can be broadly classified into two main categories: physics-driven methods \cite{yang2019improving} and data-driven methods \cite{horvath2020bayesian, guemas2016review}. Recent surveys have also provided comprehensive reviews of these data-driven spatiotemporal prediction methodologies \cite{survey2025, survey2022}. Physics-driven approaches rely on numerical models that simulate sea ice behavior by solving coupled dynamic and thermodynamic equations. These models incorporate physical laws and external forcing factors such as wind, ocean currents, and atmospheric conditions to predict sea ice changes. While they provide interpretable and physically consistent forecasts, their accuracy is often limited by model complexity, parameterization uncertainty, and high computational costs.

In contrast, data-driven approaches (particularly those based on deep learning) have shown significant improvement in Arctic SIC prediction due to their ability to automatically extract spatiotemporal features and nonlinear patterns from large volumes of observational data. Deep learning models such as IceNet \cite{icenet21}, SICNet \cite{ren2022data}, FCNet \cite{fcnet25}, and IceDiff \cite{icediff25} have demonstrated effective performance in capturing the intricate relationships among multiple variables, improving prediction accuracy. These advances highlight the growing potential of deep learning in addressing the challenges of SIC forecasting in a rapidly changing Arctic environment.

Despite these successes, many deep learning architectures, particularly those based on Transformers \cite{NIPS2017_transformer}, encounter computational bottlenecks when processing the extensive spatiotemporal data inherent in climate modeling. This challenge has spurred research into alternative architectures with greater efficiency for long-sequence modeling.

Recently, State Space Models (SSMs) have shown their superiority in natural language processing \cite{mamba24cikm}, image restoration \cite{zhou25cvpr, li25cvpr, zzh25jstars}, and hyperspectral image processing \cite{xyc25grsl, llw25tgrs, llh25tgrs}, particularly due to their linear complexity when dealing with long sequences. By formalizing discrete state space equations recursively, Mamba can capture long-range dependencies \cite{mamba24}, thereby improving sequence prediction quality in abrupt climatic impacts. Additionally, the linear-time complexity of Mamba in processing sequential data makes it particularly suitable for handling massive remote sensing satellite data, effectively reducing computational overhead while maintaining prediction accuracy.

However, adapting Mamba-based models to SIC prediction faces challenges due to two main limitations:  \textbf{(1) \textit{It emphasizes the spatial continuous information in a single image, but struggles to effectively model temporal correlations.}} It emphasizes the spatial continuous information in a single image, but struggles to effectively model spatiotemporal correlations due to the naive data processing strategy. As shown in Fig. 1(a), this baseline approach recursively processes frames that are flattened into 1D sequences. Unlike Convolutional RNNs (e.g., ConvLSTM) that preserve 2D spatial structures while iterating through time, the standard flattening process in vanilla Mamba converts the 3D spatiotemporal volume into a global 1D sequence using a fixed scanning order (e.g., raster scan). This method creates a significant theoretical gap between the 1D scanned sequence and the inherent 3D sea ice dynamics: spatially and temporally adjacent grid points in the real world become widely separated in the flattened 1D sequence. For instance, the index distance between a pixel at time $t$ and the same pixel at time $t + 1$ equals the total number of pixels in a frame. This disruption of spatiotemporal locality makes it difficult for the SSM to capture the local evolution patterns of sea ice changes effectively. Therefore, how to effectively model the temporal correlations among sea ice sequences poses a critical challenge. \textbf{(2) \textit{The Mamba model is prone to missing boundary details of the Arctic margin regions.}} Mamba performs its attention mechanism via a scanning process, which results in the erosion of intrinsic local spatial details. For the Arctic sea ice margin, where boundary details are often subtle and scattered across the image, existing Mamba-based methods struggle to retain the critical spatial cues. Hence, how to seamlessly incorporate these spatial details while concurrently improving prediction performance is a pivotal challenge.

To address these limitations, as shown in Fig. \ref{fig_frame}, we propose the \textbf{F}requency-Enhanced \textbf{H}ilbert Scanning \textbf{Mamba} framework (\textbf{FH-Mamba}), which aims to fully exploit the temporal and local features for short-term Arctic sea ice prediction. Firstly, to mitigate the limitation of temporal correlation, we introduce \textit{3D Hilbert scan mechanism} that capitalizes on the inherent locality of the Hilbert curve to enhance the spatio-temporal learning capabilities of the vanilla Mamba. The differences between vanilla Mamba and 3D Hilbert scanning are depicted in Fig. \ref{fig:mot}. The scanning is performed across both temporal and spatial dimensions, enhancing the model’s ability to capture temporal-level local information. Secondly, to address the loss of spatial detail in marginal regions, we use wavelet transform to enhance the high-frequency details. To further capture the complex dependencies across the sequence and frequency features, we design \textit{Hybrid Shuffle Attention (HSA)} module to aggregate the sequence and frequency features by calculating attentions within corresponding channels. As illustrated in Fig. \ref{fig:mot}(c), this design effectively leverages the complementary information between sequence and frequency domains, leading to improved spatial detail preservation in marginal regions.

Our main contributions can be summarized as follows:

\begin{itemize}

\item We propose FH-Mamba, a novel framework that employs 3D Hilbert scanning mechanism for Arctic SIC prediction. The scanning is performed across both temporal and spatial dimensions,  enhancing the model’s capability to capture temporal-level local information. 

\item We employ wavelet transform to enhance high-frequency details in the margin regions. Additionally, we design a Hybrid Shuffle Attention (HSA) module to fuse sequence and frequency features, effectively exploiting their complementary information.

\item We conduct extensive experiments on two benchmark datasets demonstrating that our FH-Mamba outperforms state-of-the-art methods. To support the Arctic and climate research community, we will release the code publicly.
 
\end{itemize}

\section{Related Works}

\subsection{Deep Learning-based Sea Ice Prediction}

Deep learning has made remarkable strides in Arctic sea ice concentration (SIC) prediction by leveraging deep neural networks to model the complex, nonlinear relationships inherent in observational data. For instance, ConvLSTM is employed to model the spatio-temporal correlations for SIC prediction \cite{chi2017prediction}, demonstrating superiority over traditional autoregressive models. To better incorporate spatial context, subsequent works utilized architectures like Convolutional Neural Networks (CNNs) \cite{kim2020prediction} and Transformer \cite{sicnet25gmd} for weekly SIC prediction. More recently, U-Net \cite{unet2015CoRR} based models have become prominent, such as SICNet \cite{ren2022data} and FCNet \cite{fcnet25} for short-term forecasts. IceDiff \cite{icediff25} combines Swin Transformer \cite{swintransformer2021} with a diffusion model in a two-stage framework for Arctic SIC prediction and super-resolution. Recently, IceMamba \cite{Wang2025} employs SSMs for Pan-Arctic sea ice prediction. 

Despite their effectiveness, Mamba-based SIC prediction methods do not fully exploit the temporal correlations among SIC frames. In this paper, we employ a 3D Hilbert mechanism to enhance the temporal learning capabilities. Employing the 3D Hilbert mechanism enhances temporal learning by preserving spatial and temporal locality, enabling the model to better capture sequential dependencies and fine-grained spatiotemporal patterns in Arctic data.

\begin{figure*}[]
\centering
\includegraphics[width=0.8\textwidth]{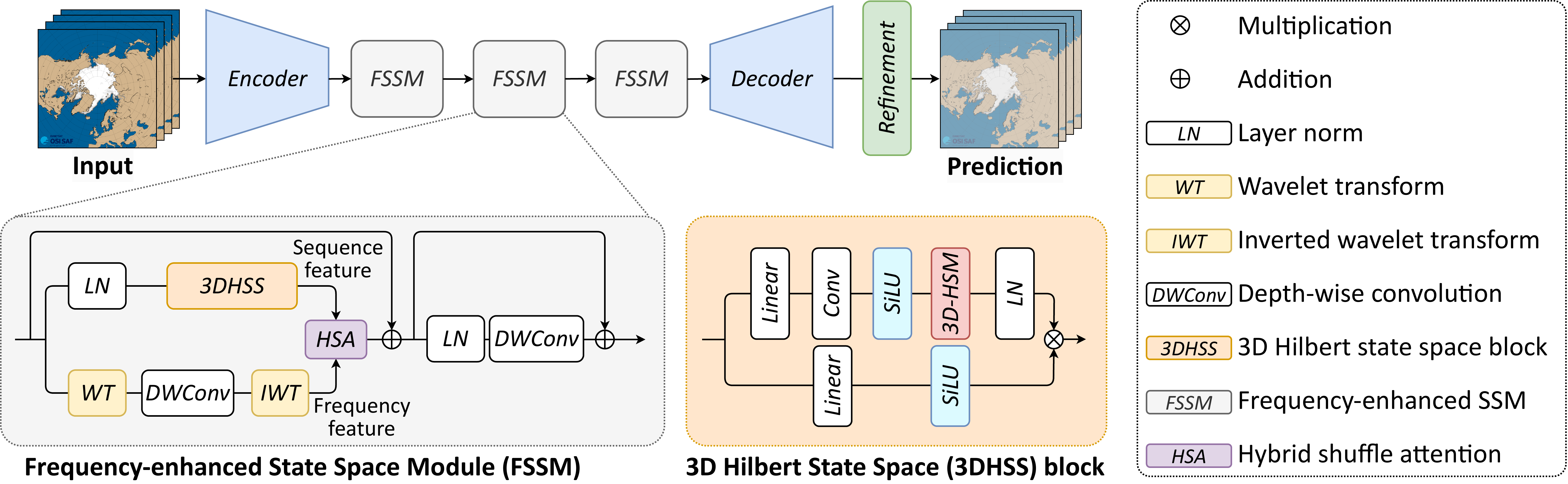}
\caption{Framework of our Frequency-Enhanced Hilbert scanning Mamba (\textbf{FH-Mamba}) for short-term Arctic sea ice concentration prediction. It is composed of a feature encoder, a series of Frequency-enhanced State Space Module (FSSM), and a feature decoder. The FSSM employs the 3D Hilbert State Space (3DHSS) block and wavelet transform to capture spatio-temporal feature dependencies. In 3DHSS, we use 3D Hilbert scanning mechanism to capture temporal-level local information. Furthermore, Hybrid Shuffle Attention (HSA) module is designed to fuse sequence and frequency features, effectively exploiting their complementary information.}
\label{fig_frame}
\end{figure*}

\subsection{State Space Models for Vision Task}

SSMs, particularly Mamba-based architectures, are being actively adapted for computer vision tasks, leveraging their efficiency in modeling long-range dependencies beyond 1D sequences. To better handle 2D spatial information, DefMamba \cite{defmamba2025CVPR} introduces deformable mamba blocks that dynamically adjust the scanning path to focus on salient image features. GroupMamba \cite{GroupMamba2025CVPR} addresses scaling challenges by proposing modulated group Mamba layer, which processes channel groups with independent, multi-directional scanning to improve stability and performance. Hybrid approaches have also emerged, such as MambaVision \cite{mambavision2025CVPR}, which integrates self-attention blocks into the final layers of a Mamba backbone to enhance its capacity for capturing long-range spatial relationships. For deployment on resource-constrained devices, EfficientViM \cite{efficientVIM2025CVPR} presents an architecture that reduces computational cost by performing channel mixing within compressed hidden states. To enhance the temporal feature learning, RainMamba \cite{rainmamba24} uses Hilbert scanning to exploit sequence correlations in videos for deraining task. 

Unlike previous works, we employ wavelet transform to enhance fine-grained details in the Arctic margin regions. Furthermore, we design a HSA module to effectively exploit the complementary information between sequence and frequency features. This enables the model to capture more comprehensive spatiotemporal dependencies and improves prediction accuracy.

\section{Methodology}

\subsection{Problem Statement}

We address the task of short-term Arctic sea ice concentration (SIC) prediction, which aims to forecast future daily SIC maps based on a sequence of past observations. Specifically, given the SIC data from the past $L_i$ days, represented as a sequence $\mathbf{X} = \{x_l\}_{l=1}^{L_i} \in \mathbb{R}^{L_i \times C \times H \times W}$, where $L_i$ denotes the input sequence length, the goal is to model the conditional distribution $p(\mathbf{Y}|\mathbf{X})$ of the next $L_o$ frames, denoted as $\mathbf{Y} = \{y_l\}_{l=1}^{L_o} \in \mathbb{R}^{L_o \times C \times H \times W}$, where $L_o$ represents the output prediction length. Here $H$ and $W$ refer to the spatial resolution (height and width) of each SIC frame, and $C = 1$ indicates that only the SIC variable is used in the prediction. This formulation treats sea ice prediction as a spatiotemporal sequence forecasting problem, where the model must learn to capture both spatial structures and temporal dynamics in the evolving sea ice field. Accurately modeling $p(\mathbf{Y}|\mathbf{X})$ is critical for anticipating short-term Arctic sea ice behavior.

\subsection{Overall Framework}

The proposed FH-Mamba framework is shown Fig. \ref{fig_frame}, which is composed of a feature encoder, a series of Frequency-enhanced State Space Module (FSSM), and a feature decoder. The encoder contains several convolutional blocks, each consisting of two-dimensional (2D) convolution, layer normalization, and LeakyReLU activation, to extract spatial features. The obtained spatial features are successively fed into FSSM for spatial and temporal feature exploitation. After that, a decoder containing several transposed convolutional blocks, which include 2D transposed convolution, group normalization, and LeakyReLU activation, is employed to reconstruct the ground truth frames. It should be noted that two depth-wise convolution layers are employed as the refinement block to optimize the spatial details of the SIC prediction results.

The FSSM is the critical component in our FH-Mamba, and details of the module are shown in Fig. \ref{fig_frame}. The input features are fed into two branches, namely Mamba and frequency branches. The Mamba branch uses 3D Hilbert State Space (3DHSS) block to exploit the spatial-temporal features. The frequency branch employs wavelet to enhance the high-frequency details. The output of both branches are fused via Hybrid Shuffle Attention (HSA). Afterwards, a depth-wise convolution layer is employed for non-linear feature transformation. Next, we introduce the 3D Hilbert scanning mechanism in Section III. C, elaborate the 3DHSS block in Section III. D, and present the Hybrid Shuffle Attention in Section III. E. The overall optimization objective is summarized in Section III. F.

\subsection{3D Hilbert Scanning Mechanism}

The Hilbert curve is a continuous space-filling curve that maps a one-dimensional interval onto a multi-dimensional space while preserving locality \cite{chen21esa}. Originally proposed by David Hilbert \cite{hilbert1891ueber}, the curve has found extensive use in computer graphics \cite{keller22}, spatial indexing \cite{uddin18}, and deep learning \cite{liang24pointmamba} due to its ability to maintain spatial proximity between neighboring points.

\textbf{3D Hilbert Curve.}  Let $H:[0,1]\rightarrow [0,1]^d$ be a $d$-dimensional Hilbert curve of order $n$, where $d$ is the spatial dimensionality and $n$ controls the resolution. In discrete form, a Hilbert curve of order $n$ in $d$ dimensions maps a scalar index $i\in\{0,1,\ldots, 2^{dn}-1\}$ to a coordinate $H(i)\in\{0,1,\ldots, 2^n-1\}^d$. The mapping ensures that consecutive indices correspond to spatially adjacent or nearby points, thereby preserving locality \cite{moon01tkde, butz71tc}. In the context of 3D data (the Arctic sea ice concentration spatio-temporal volumes), the 3D Hilbert curve extends the locality-preserving property into three-dimensional space \cite{cjf22tip}. It defines a traversal path through a 3D cube of size $2^n\times 2^n \times 2^n$, such that adjacent indices in the one-dimensional sequence correspond to neighboring voxels in the 3D grid. This is particularly valuable in SSMs benefits from spatial or temporal coherence. Let $V\in \mathbb{R}^{T\times C\times H\times W}$ be a 3D spatio-temporal volume, where $T$ is the temporal length, $C$ is the number of channels, and $H\times W$ is the spatial resolution. The 3D Hilbert scanning function $\phi: \mathbb{Z}^4\rightarrow \mathbb{Z}$ maps each voxel at position $(t, c, h, w)$ to a unique index $i$ along the Hilbert curve, producing a flattened sequence $\{v_i\}^N_{i=1}$, where $N=T\cdot C\cdot H\cdot W$. This scan preserves the spatial and temporal proximity, allowing for effective modeling of localized correlations in the Arctic Sea Ice Concentration (SIC) data sequence. In this paper, we use the Generalized Hilbert algorithm for space filling for 3D data. The idea is to recursively apply the following template to obtain a Hilbert-like curve. A general rectangle with a known orientation is split into three regions (``up", ``right", ``down"), for which the function calls itself recursively, until a trivial path can be produced \cite{wj18tvcg}.

\begin{figure}
    \centering
    \includegraphics[width=0.7\linewidth]{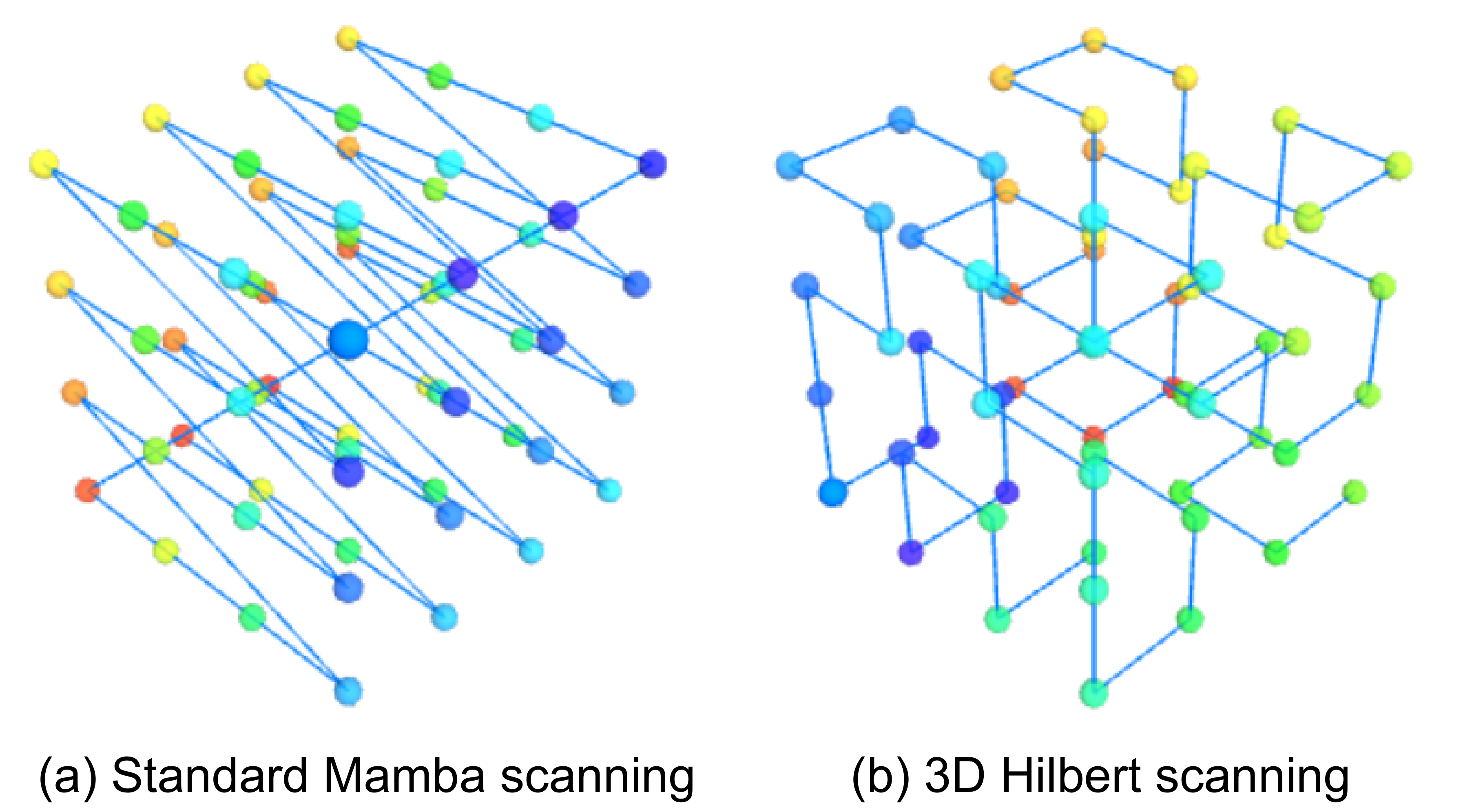}
    \caption{Trajectories of standard Mamba and 3D Hilbert scanning.}
    \label{fig:trajectory}
\end{figure}

\textbf{Comparisons of 3D Hilbert Scanning and Standard Mamba Scanning.} Standard Mamba scanning typically flattens the spatio-temporal input data into a 1D sequence in a linear order. This flattening disrupts the intrinsic spatial and temporal adjacency, leading to a loss of structural information crucial for capturing ice dynamics. However, 3D Hilbert scanning preserves local neighborhood continuity in both spatial and temporal dimensions, maintaining the inherent topology of Arctic sea ice evolution. For prediction accuracy, standard Mamba scanning suffers from limited spatial detail reconstruction, especially in marginal ice zones where precise prediction is crucial. At the same time, 3D Hilbert Scanning improves spatial detail preservation by enhancing the sequence's locality, leading to better reconstruction of complex sea ice patterns and reduced prediction bias in marginal areas. Fig. \ref{fig:trajectory} illustrates the difference between Standard Mamba scanning and 3D Hilbert scanning in terms of their traversal paths through a 3D spatiotemporal volume. Standard Mamba scanning visits points along a fixed direction (e.g., time-major or spatial-major order). This approach does not preserve the spatial or temporal locality; neighboring points in 3D space may be far apart in the flattened sequence. However, 3D Hilbert scanning traverses the 3D volume in a way that maintains local continuity. The scanning path visits neighboring points in a spatially and temporally coherent order, preserving locality and structural integrity of the input data. It is evident that 3D Hilbert scanning (right) provides a more topology-aware and locality-preserving trajectory compared to the standard Mamba scanning strategy (left), which is essential for modeling complex spatial-temporal patterns in Arctic sea ice prediction.

\textbf{Hilbert Scanning vs. Graph Approaches.} It is worth noting that while 3D Hilbert scanning preserves locality better than raster scanning, it still faces the boundary problem,' where spatially adjacent voxels may be distant in the flattened 1D sequence. Graph Neural Networks (GNNs) could theoretically solve this by explicitly modeling adjacency via edges. However, for dense regular grid data like sea ice data, GNNs often incur prohibitive computational costs due to the massive number of nodes and edges required for fine-grained prediction. In contrast, our Mamba-based approach maintains linear complexity. Furthermore, the limitation of the Hilbert curve is effectively mitigated in FH-Mamba by two factors: First, the SSM's inherent ability to model long-range dependencies allows it to connect distant elements in the sequence. Second, the CNN-based encoder and the parallel Wavelet branch capture 2D spatial structures and high-frequency boundaries independently of the scanning order, ensuring that local adjacency information is not lost due to sequence discontinuity.

\subsection{3D Hilbert State Space Block}

\textbf{Mamba Block.} For the Mamba block, we adopt a structure similar to that of Vision Mamba \cite{vim24icml}. As shown in Fig. \ref{fig_frame}, it consists of LayerNorm (LN), Linear layers, 1D Convolution, and 3D Hilbert SSM. Given an input $\mathbf{X}\in \mathbb{R}^{B\times L\times D}$, where $B$, $L$, and $D$ denote the batch size, the length of 1D data, and the feature dimension, respectively. The output $\mathbf{X}'$ is computed as:
\begin{equation}
    \mathbf{X}'_1= \textrm{LN}(\mathbf{X}),
\end{equation}
\begin{equation}
    \mathbf{X}'_2= \textrm{3D\_Hilbert\_SSM}(\textrm{Silu}(\textrm{Conv}(\textrm{Linear}(\mathbf{V}'_1)))),
\end{equation}
\begin{equation}
    \mathbf{V}'=\textrm{Linear}(\mathbf{V}'_2 \odot \textrm{Silu}(\textrm{Linear}(\mathbf{V}'_1))),
\end{equation}
where $\odot$ denotes the element-wise multiplication. It is worth noting that we use a 3D Hilbert scan instead of the original global scan in Vision Mamba. Both forward and backward scans are employed in our implementation.

\begin{figure*}[]
\centering
\includegraphics[width=0.8\linewidth]{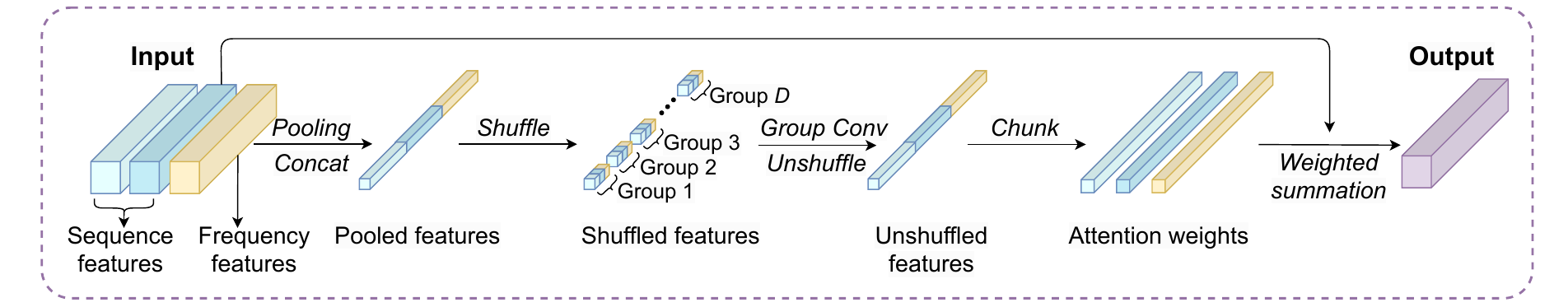}
\caption{Illustration of the Hybrid Shuffle Attention (HSA). The input sequence and frequency features $\{\mathbf{x}^1, \mathbf{x}^2, \mathbf{x}^f\}$ are handled by pooling and concatenation to generate $\hat{\mathbf{X}}$. It undergoes the shuffle operation and results in $\hat{\mathbf{X}}'$. Afterwards, $\hat{\mathbf{X}}'$ are split into $D$ groups. Group convolution and unshuffle operation are employed, generating unshuffled weights $\hat{\mathbf{A}}$. The weights are further chunked and reshaped into attention weights $\{\mathbf{A}^1, \mathbf{A}^2, \mathbf{A}^f\}$. Finally, the output is computed by performing weighted summation of the input features  and the attention weights.}
\label{fig:hsa}
\end{figure*}

\subsection{Hybrid Shuffle Attention}

Hybrid Shuffle Attention (HSA) aggregates the 3D Hilbert sequence and frequency features by calculating attentions within corresponding channels. This design enables it to capture complex dependencies across distinct sequences, thus better leveraging complementary information from different scanning directions. As shown in Fig. \ref{fig:hsa}, supposing two sequence features $\mathbf{x}^1$ and $\mathbf{x}^2$, and frequency features $\mathbf{x}^f$, we first apply the spatial average pooling to reduce the computational cost, and then concatenate as:

\begin{equation}
\begin{aligned}
\hat{\mathbf{X}} &=\textrm{concat}(\textrm{GAP}(\mathbf{x}^1, ~ \mathbf{x}^2, ~ \mathbf{x}^f)) \\
&= [x^1_1, \cdots, x^1_D, ~ 
x^2_1, \cdots, x^2_D, ~ 
x^f_1, \cdots, x^f_D],
\end{aligned}
\end{equation}
where $x^k_d$ is the pooled feature in $d$-th channel of $k$-th sequence, and $x^f_d$ is the pooled feature in $d$-th channel of the frequency feature. $D$ is the feature dimension. Afterwards, we employ sequence shuffle operation to rearrange feature as follows:

\begin{equation}
\begin{aligned}
\hat{\mathbf{X}}' &= \textrm{shuffle}(\hat{\mathbf{X}})\\
&= [x^1_1, x^2_1, x^f_1, ~ x^1_2, x^2_2, x^f_2, ~ \cdots, ~
x^1_D, x^2_D, x^f_D].
\end{aligned}
\end{equation}

Afterwards, we employ group convolution with group size three to obtain the channel-wise attention weights and unshuffle the weights back to the original order as follows:
\begin{equation}
\begin{aligned}
\hat{\mathbf{A}} &= \textrm{unshuffle}(\textrm{GConv}(\hat{\mathbf{X}}')) \\
&=[w^1_1, \cdots, w^1_D, ~ w^2_1, \cdots, w^2_D, ~
w^f_1, \cdots, w^f_D],
\end{aligned}
\end{equation}
where GConv($\cdot$) is the group convolution. The unshuffled weights $\hat{\mathbf{A}}$ are chunked as $\{\mathbf{A}^1, \mathbf{A}^2, \mathbf{A}^f\}=\textrm{chunk}(\hat{\mathbf{A}})$, where chunk($\cdot$) is the chunk operation. Finally, weight summation is used to generate the output, which can be formulated as:
\begin{equation}
\mathbf{Y}=\mathbf{A}^1 \mathbf{x}^1+\mathbf{A}^2 \mathbf{x}^2+\mathbf{A}^f \mathbf{x}^f,
\end{equation}
and $\mathbf{Y}$ is the output of HSA.

\subsection{Loss Function}

To ensure that the predicted SIC sequences aligns with the ground truth at the pixel level, we use L1 loss to measure the discrepancy between the final prediction $\hat{\mathbf{Y}}$ and the ground truth target $\mathbf{Y}$:
\begin{equation}
    \mathcal{L}_{rec}= ||\hat{\mathbf{Y}}-\mathbf{Y}||_1.
\end{equation}

We also use the gradient loss to preserve edges and fine details by comparing gradients:
\begin{equation}
    \mathcal{L}_{grad}=||\nabla\hat{\mathbf{Y}}-\nabla\mathbf{Y}||_1,
\end{equation}
where $\nabla$ denotes the gradient operator. The final total loss is a weighted summation of both loss terms:
\begin{equation}
    \mathcal{L}_{total}=\mathcal{L}_{rec}+\lambda\mathcal{L}_{grad},
\end{equation}
where $\lambda$ is a hyperparameter that balances the reconstruction and detail preservation.

\section{Experimental Results and Analysis}

\subsection{Datasets}

We evaluate our proposed FH-Mamba model for Sea Ice Concentration (SIC) prediction using the OSI-450a1 dataset. As an updated version of the OSI-450a1 dataset \cite{lavergne2019version}, OSI-450a1 is sourced from the OSI-SAF website and represents a full reprocessing of SIC data from 1978 to 2020 with optimized data processing. It includes coarse-resolution images ($432\times 432$ pixels) from SMMR (1978-1987), SSM/I (1987-2008), and SSMIS (2006-2020) satellites, along with ECMWF ERA-Interim data \cite{dee2011era}. Model assessments are performed within the Arctic Monitoring and Assessment Program (AMAP) region \cite{hung2010atmospheric}, north of the Arctic Circle, to demonstrate the significance of SIC prediction for understanding the sea ice system's impact on environmental and human activities. We utilize data spanning from October 25, 1978, to December 31, 2020, dividing it into a training set (1978–2010), a validation set (2011–2015), and a test set (2016–2020), containing 11,729, 1,799, and 1,800 samples, respectively.

To further evaluate the generalization capability of our model, especially on data with different sensor characteristics and significantly higher spatial resolution, we also conduct experiments on the AMSR2 Sea Ice Concentration (SIC) dataset \cite{Spreen2008SeaIR}. This dataset is provided by the University of Bremen, sourced from the AMSR2 sensor aboard the Japan Aerospace Exploration Agency (JAXA)'s GCOM-W1 satellite. It offers daily sea ice concentration data from July 2012 to the present with a spatial resolution of 3.125 km, which is substantially finer than the 25 km resolution of the OSI-450a1 dataset. We employed the Spatio-Temporal Inverse Distance Weighting algorithm with a Gaussian kernel to interpolate the missing data points in AMSR2 data. Subsequently, pixels with more than 95\% missing data throughout the time series were identified as land and excluded from the analysis. For the AMSR2 dataset, we use the data from August 12, 2012, to December 31, 2023, serving as the training set. Data from January 1, 2024, to December 31, 2024, are employed as the validation set and data from January 1, 2025, to August 31, 2025, are employed as the test set, containing 4,102, 339, and 188 samples, respectively.

\subsection{Experimental Setup}

To ensure temporal continuity and data consistency, we performed specific preprocessing prior to input generation. Since the OSI-450a1 dataset is a reprocessed product without spatial gaps (e.g., pole holes), we focused on filling missing dates using the mean of the preceding and succeeding valid frames. Additionally, all land pixels were initialized to zero. Input/output samples are generated using a 14-day sliding window over the SIC time series. Specifically, the SIC data from the past 14 days are used to predict the SIC data for the following 14 days. FH-Mamba is trained using the AdamW optimizer with an initial learning rate of 0.001 and a batch size of 4. Model performance is quantified using Root Mean Square Error (RMSE), Mean Absolute Error (MAE), and Nash-Sutcliffe Efficiency (NSE). For RMSE and MAE, lower values indicate better prediction performance, while for NSE, values closer to 1 reflect better model performance. The experiments were conducted on hardware with an AMD EPYC 9754 CPU and a GeForce RTX 4090 GPU. Additionally, an early stopping mechanism is employed to prevent overfitting: the training process is terminated if the MAE value on the validation set fails to improve for 10 consecutive epochs.

To ensure a fair and direct comparison, all baseline results reported in this study were obtained by retraining the models using their official source code on our OSI-450a1 and AMSR2 dataset. We maintained a consistent experimental setup across all models, including identical data splits, batch sizes, and learning rates. However, to allow each baseline to achieve its optimal performance as intended by its authors, we adopted the specific loss functions and optimizers proposed in their respective original publications. This approach prevents potential performance degradation that could arise from using non-native optimization schemes and ensures that our comparison is against each model's best possible capability within our experimental framework.

\begin{table*}[ht]
\centering
\caption{Prediction performance of different methods on the OSI-450-a1 dataset. The \textbf{bold} and \underline{underline} denote the best and second results. (NSE values adjusted for consistency with RMSE)}
\setlength{\tabcolsep}{4pt}
\begin{tabular}{c|ccccc|ccccc|ccccc}
\toprule
\multirow{2}{*}{\textbf{Method}} & \multicolumn{5}{c|}{\textbf{RMSE (\%)} $\downarrow$}             & \multicolumn{5}{c|}{\textbf{MAE (\%)} $\downarrow$}              & \multicolumn{5}{c}{\textbf{NSE (\%)} $\uparrow$}      \\ \cmidrule{2-16}
 & 2016   & 2017   & 2018   & 2019   & 2020   & 2016   & 2017   & 2018   & 2019   & 2020   & 2016   & 2017   & 2018   & 2019   & 2020    \\ \midrule
ConvLSTM  & 9.148  & 8.466  & 8.838  & 8.738  & 9.093  & 2.967  & 2.733  & 2.755  & 2.749  & 2.855  & 92.597 & 94.537 & 94.223 & 94.083 & 93.069 \\ 
PredRNN   & 9.054  & 8.356  & 8.505  & 8.360  & 8.827  & 2.938  & 2.679  & 2.642  & 2.635  & 2.777  & 92.691 & 94.729 & 94.636 & 94.674 & 93.431 \\ 
PredRNNv2 & 8.807  & 8.219  & 8.526  & 8.362  & 8.649  & 2.852  & 2.665  & 2.671  & 2.673  & 2.750  & 93.196 & 94.967 & 94.746 & 94.753 & 93.844 \\
IceNet    & 9.511  & 8.636  & 9.088  & 8.803  & 9.128  & 3.064  & 2.715  & 2.827  & 2.714  & 2.819  & 91.878 & 94.207 & 93.546 & 93.939 & 92.577 \\
SimVP     & 7.472  & 6.862  & 7.089  & \underline{6.720}  & 7.091  & 2.408  & 2.308  & 2.210  & \underline{2.095}  & 2.233  & 95.296 & 96.578 & 96.331 & \underline{96.698} & 95.870 \\
SwinLSTM  & 7.258  & 6.769  & \underline{6.784} & 6.739  & \underline{6.881} & 2.341  & 2.211  & \underline{2.160} & 2.167  & 2.176  & 95.508 & 96.652 & \underline{96.643} & 96.501 & \underline{96.107} \\ 
VMRNN     & 7.224  & 6.901  & 7.089  & 6.901  & 7.101  & 2.418  & 2.268  & 2.282  & 2.207  & 2.285  & 95.683 & 96.462 & 96.276 & 96.429 & 95.975 \\
FCNet     & 7.276  & 6.751  & 6.884  & 6.958  & 6.897  & \underline{2.327}  & \underline{2.131}  & 2.193  & 2.128  & \underline{2.168}  & 95.671 & 96.573 & 96.543 & 96.691 & 96.099 \\
IceMamba  & \underline{7.135}  & \underline{6.676}  & 6.850  & 6.735  & 6.885  & 2.424  & 2.258  & 2.230  & 2.195  & 2.243  & \underline{95.745} & \underline{96.734} & 96.577 & 96.538 & 96.105 \\

\textbf{FH-Mamba (Ours)} & \textbf{7.069} & \textbf{6.606} & \textbf{6.777} & \textbf{6.602} & \textbf{6.798} & \textbf{2.211} & \textbf{2.051} & \textbf{2.049} & \textbf{1.994} & \textbf{2.074} & \textbf{95.849} & \textbf{96.825} & \textbf{96.650} & \textbf{96.767} & \textbf{96.152} \\ \bottomrule
\end{tabular}
\label{tab:comparation}
\end{table*}

\begin{table}[htbp]
\centering
\caption{Statistical significance analysis (paired t-test) of RMSE results between fh-mamba and baseline models (2016--2020).}
\label{tab:significance}
\begin{tabular}{lccc}
\toprule
\textbf{Comparison} & \textbf{t-statistic} & \textbf{p-value} & \textbf{Significance} \\ \midrule
FH-Mamba vs. SimVP    & 5.9557  & 0.0040 & ** ($p < 0.01$) \\
FH-Mamba vs. SwinLSTM & 3.5780  & 0.0232 & * ($p < 0.05$) \\
FH-Mamba vs. VMRNN    & 9.2211  & 0.0008 & *** ($p < 0.001$) \\
FH-Mamba vs. FCNet    & 3.8632  & 0.0181 & * ($p < 0.05$) \\
\bottomrule
\end{tabular}
\end{table}

\begin{table}[h]
\centering
\caption{Performance comparison of different methods on the AMSR2 dataset. The \textbf{bold} and \underline{underline} denote the best and second results.}
\begin{tabular}{c|ccc}
\toprule
\textbf{Model} & \textbf{MAE (\%)} $\downarrow$ & \textbf{RMSE (\%)} $\downarrow$ & \textbf{NSE (\%)} $\uparrow$ \\ \midrule
ConvLSTM   & 3.75 & 14.32 & 88.02 \\
PredRNN    & 3.65 & 14.02 & 88.59 \\
PredRNNv2  & 3.50 & 13.51 & 89.09 \\
IceNet     & 3.84 & 14.96 & 87.75 \\
SimVP      & 2.89 & 10.85 & 91.58 \\
SwinLSTM   & 2.74 & 10.58 & 91.69 \\
VMRNN      & 2.65 & \underline{10.16} & \underline{91.91} \\ 
FCNet      & \underline{2.63} & 10.21 & 91.89 \\
IceMamba   & 2.69 & 10.42 & 91.81 \\
\textbf{FH-Mamba (Ours)}   & \textbf{2.54} & \textbf{10.05} & \textbf{92.04} \\ 
\bottomrule
\end{tabular}
\label{tab:model-comparison-amsr2}
\end{table}

\subsection{Comparison with State-of-the-Art Methods}

We compare FH-Mamba with 8 state-of-the-art (SOTA) methods: ConvLSTM \cite{chi2017prediction}, PredRNN \cite{wang17predrnn}, PredRNNv2 \cite{wang2022predrnn}, IceNet \cite{icenet21}, SimVP \cite{gao2022simvp}, SwinLSTM \cite{tang2023swinlstm}, VMRNN \cite{tang2024vmrnn}, FCNet  \cite{fcnet25} and IceMamba \cite{Wang2025}. ConvLSTM \cite{chi2017prediction} extends standard LSTMs by replacing full connections with convolutional structures, enabling effective modeling of spatiotemporal dependencies for data sequences. PredRNN \cite{wang17predrnn} introduces a recurrent memory architecture that flows both along temporal and spatial dimensions. PredRNNv2 \cite{wang2022predrnn} improves PredRNN with a decoupled spatiotemporal memory mechanism and a gradient highway unit, enabling deeper temporal propagation and more accurate sequence generation. IceNet \cite{icenet21} combines convolutional encoders with autoregressive temporal modeling to predict Arctic SIC at grid level with uncertainty estimation. SimVP \cite{gao2022simvp} is a simple yet effective prediction framework that decouples spatial encoding from temporal dynamics. It employs convolutional networks for spatial representation and lightweight temporal blocks for efficient sequence forecasting.  SwinLSTM \cite{tang2023swinlstm} integrates Swin Transformer blocks into an LSTM framework, leveraging hierarchical attention and shifted windows to capture long-range spatial dependencies in spatio-temporal data.  VMRNN \cite{tang2024vmrnn} introduces adaptive memory update mechanisms that dynamically adjust the receptive field over time, improving the model's flexibility for complex temporal patterns. FCNet \cite{fcnet25} enhances Arctic SIC prediction by integrating adaptive frequency filtering and convolutional feature extraction to jointly capture fine-grained edges and long-term spatiotemporal variations. IceMamba \cite{Wang2025} is a state-space-based framework for seasonal Pan-Arctic sea ice forecasting. It employs Residual Selective State Space Blocks (RESSB) to capture complex spatiotemporal dependencies while maintaining linear computational complexity for long-sequence climate data.

\textbf{Results on the OSI-450-a1 dataset.} Table \ref{tab:comparation} presents the prediction performance of various methods on the OSI-450-a1 dataset from 2016 to 2020. Overall, our FH-Mamba achieves the best performance in all cases. Specifically, FH-Mamba attains the lowest RMSE values (6.60 \%–7.07 \%) and MAE values (1.99 \%–2.21 \%), significantly outperforming classical recurrent networks such as ConvLSTM, PredRNN, and PredRNNv2, as well as recent Transformer-based or frequency-based models including SwinLSTM, VMRNN, and FCNet. Meanwhile, FH-Mamba consistently yields the highest NSE scores, indicating  improved consistency in reproducing sea-ice dynamics. These results demonstrate that FH-Mamba effectively captures both short-term variations and long-range dependencies in Arctic sea-ice evolution, delivering more accurate and robust spatiotemporal forecasts than existing deep learning baselines. Furthermore, to evaluate whether the performance gain is statistically significant, we conduct a paired t-test based on the annual RMSE values from 2016 to 2020, as summarized in Table \ref{tab:significance}. All p-values are smaller than 0.05, with the comparison against VMRNN even reaching p $<$ 0.001. This indicates that the effective of FH-Mamba is statistically significant and robust, rather than being a result of random chance.

\textbf{Results on the AMSR2 dataset.} Table \ref{tab:model-comparison-amsr2} presents the performance comparison of various spatiotemporal prediction models on the AMSR2 dataset. Among all competitors, the proposed FH-Mamba achieves the best overall performance, obtaining the lowest MAE (2.54 \%) and RMSE (10.05 \%), as well as the highest NSE (92.04 \%). Compared with classical recurrent models such as ConvLSTM, PredRNN, and PredRNN-v2, FH-Mamba significantly reduces prediction errors and enhances numerical stability. It also surpasses advanced transformer- and frequency-based baselines (SwinLSTM, VMRNN, FCNet), demonstrating its capability to capture both local spatial structures and long-range temporal dependencies. These results verify the generalization capability of FH-Mamba across datasets, showing that the model effectively adapts to different input distributions and achieves accurate and robust Arctic sea ice forecasting.

\begin{figure*}[ht]
    \centering
    \includegraphics[width=0.65\textwidth]{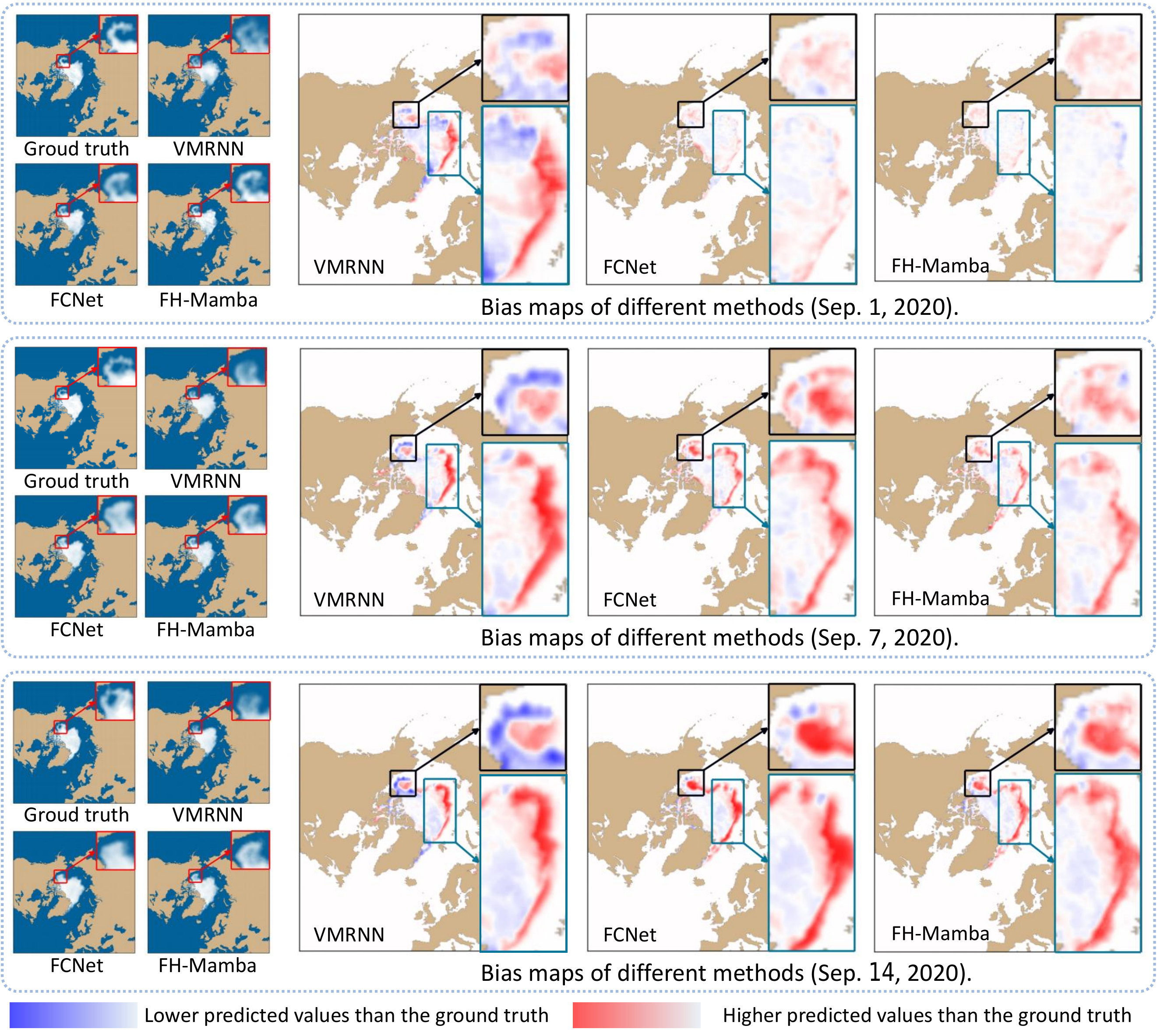}
    \caption{The visualization results of VMRNN, FCNet, and our FH-Mamba during the period from September 1 to 14, 2020. The left columns show the predicted SIC maps for each model alongside the ground truth. The right columns highlight the bias maps (prediction minus ground truth). In the bias maps, positive errors are shown in red and negative errors in blue.}
    \label{fig:vis}
\end{figure*}

\subsection{Visual and Qualitative Analysis}

\begin{figure}[ht!]
    \centering
    \includegraphics[width=3.5in]{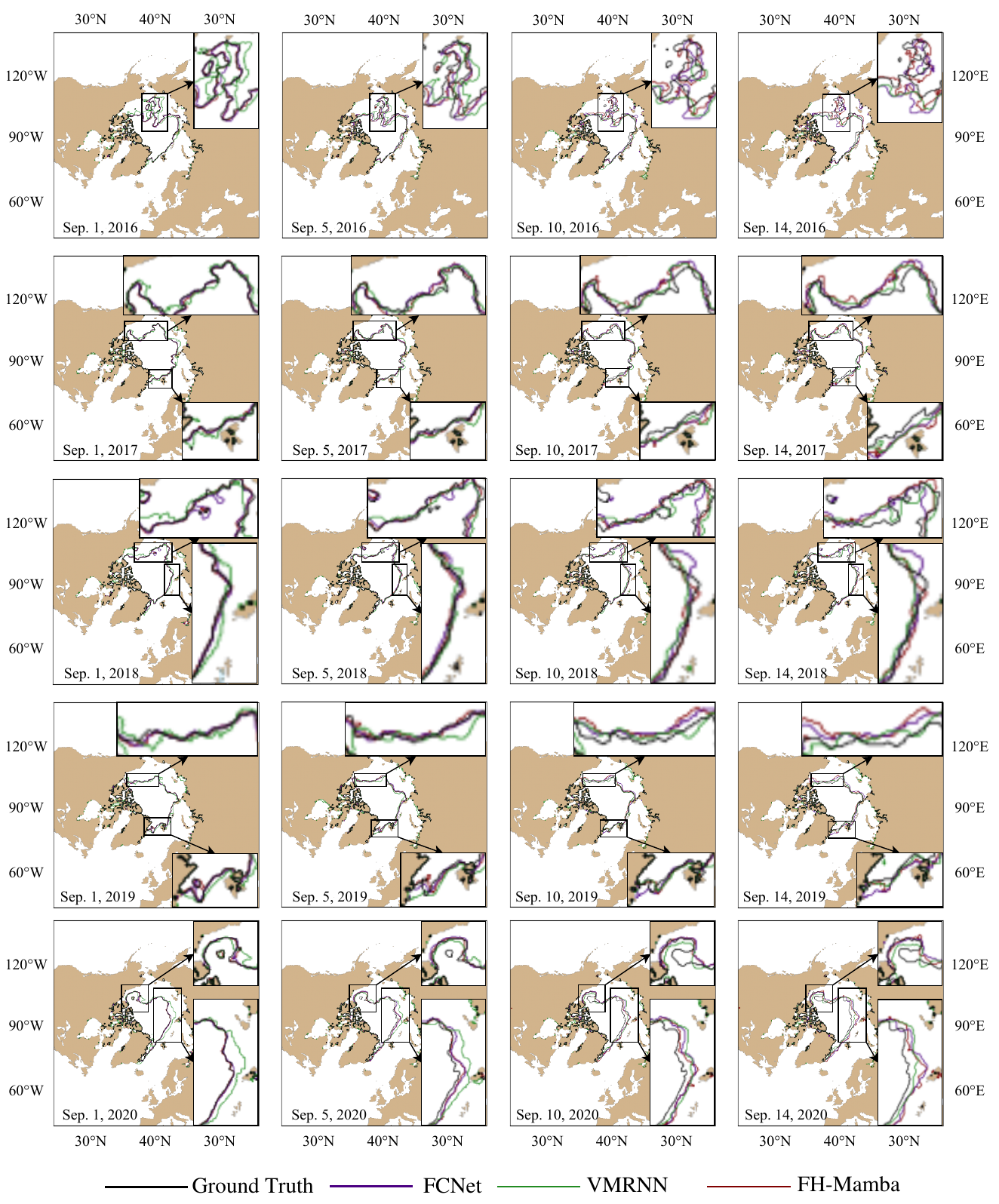}
    \caption{Comparison of SIE on September 1st, 5th, 10th, and 14th of different years using various methods. The black line represents the SIE from the ground truth map. The green line signifies the SIE forecasted by the VMRNN. The purple line denotes the SIE predicted with the FCNet. And the red line represents the SIE predicted by our FH-Mamba.}
    \label{fig:SIE1}
\end{figure}

\textbf{Visualization of Prediction.} In September 2020, Arctic sea ice reached a record‑low minimum extent, marking one of the most significant observations. This period reflects extreme Arctic sea ice loss driven by persistent warm conditions, making it a representative and challenging case for evaluating SIC prediction models \cite{stroeve2018changing}. Therefore, we compare the visualization results of our FH-Mamba with those of VMRNN and FCNet for the period from September 1 to 14, 2020, as shown in Fig. \ref{fig:vis}.  The left columns show the predicted SIC maps for each model alongside the ground truth. The right columns highlight bias maps (predicted results minus ground truth). In the bias maps, positive errors are shown in red, and negative errors in blue. As can be observed, our FH-Mamba consistently shows smaller error regions compared to VMRNN and FCNet, particularly along the Arctic margin regions, where accurate prediction is more difficult due to complex dynamics. It is evident that the wavelet branch in FSSM improves the boundary detail prediction. In addition, our FH-Mamba better preserves fine spatial structures and boundary details, especially in the zoomed-in regions. This demonstrates the capability of the model to capture local textures and edges in sea ice prediction. Furthermore, across all three dates, FH-Mamba maintains consistent performance, whereas VMRNN and FCNet show fluctuating error patterns and more significant deviations from the ground truth. These results confirm that FH-Mamba delivers more accurate and stable SIC predictions with reduced biases and improved spatial detail reconstruction, particularly in Arctic marginal ice zones.

\begin{figure}[ht!]
    \centering
    \includegraphics[width=3.5in]{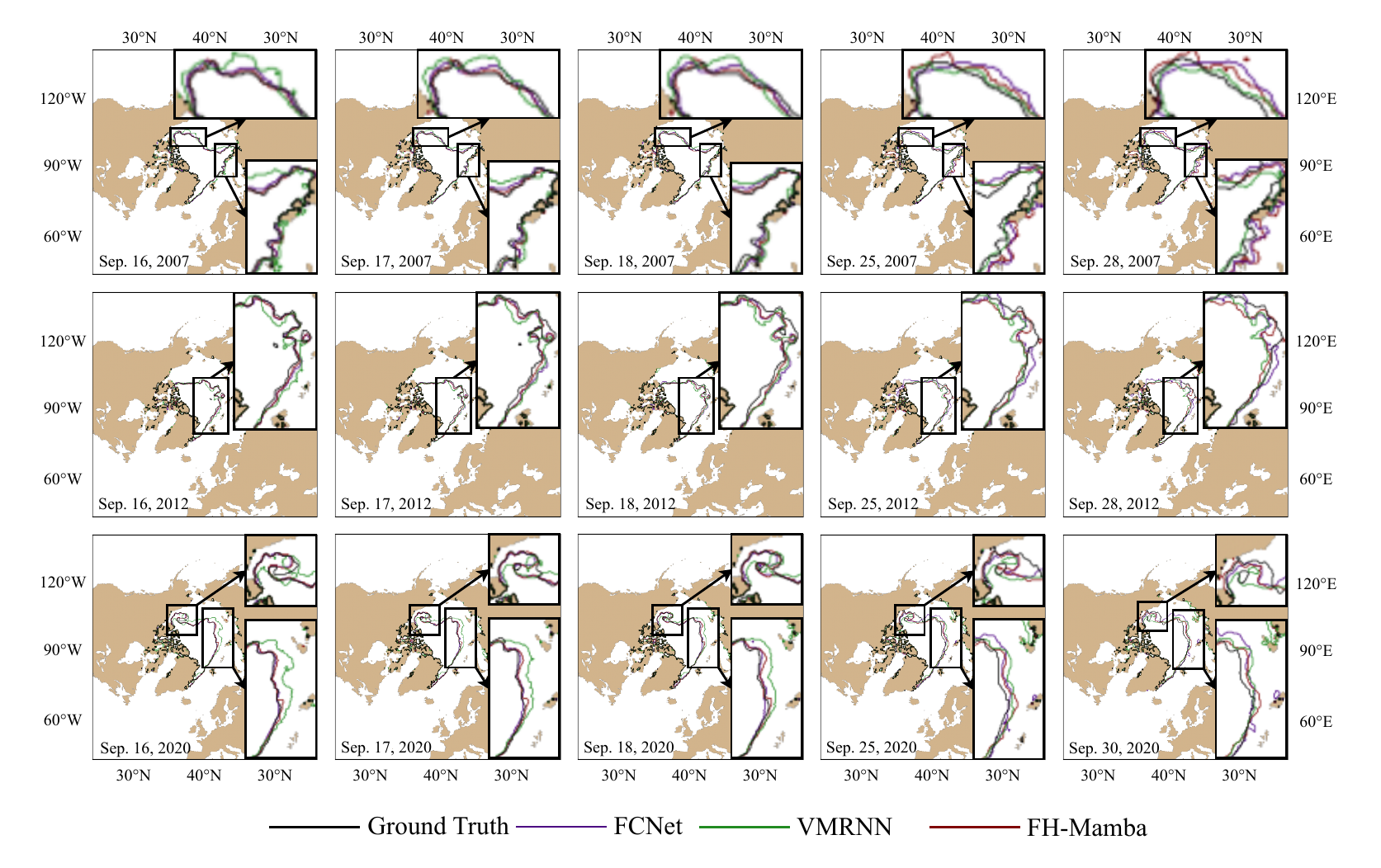}
    \caption{Among the three extreme climate events with the smallest sea ice extent on record, the Arctic sea ice coverage reached its lowest points on the following dates: September 17, 2012, with an area of 3.39 million square kilometers; followed by September 16, 2020, with an area of 3.82 million square kilometers; and September 18, 2007, with an area of 4.14 million square kilometers.}
    \label{fig:Sep9}
\end{figure}

\begin{table}[htbp]
\centering
\caption{Quantitative evaluation of Sea Ice Extent (SIE) alignment using IoU metric (2016.9--2020.9).}
\label{tab:IoU_Results}
\begin{tabular}{c|ccc}
\toprule
\textbf{Time} & \textbf{VMRNN} & \textbf{FCNet} & \textbf{FH-Mamba (Ours)} \\ \midrule
2016.9          & 0.7665         & 0.8517         & \textbf{0.8707}          \\
2017.9          & 0.8769         & 0.9058         & \textbf{0.9082}          \\
2018.9          & 0.8788         & 0.9077         & \textbf{0.9155}          \\
2019.9          & 0.8601         & 0.9022         & \textbf{0.9143}          \\
2020.9          & 0.7970         & 0.8603         & \textbf{0.8632}          \\ \midrule
\textbf{Average} & 0.8359      & 0.8855         & \textbf{0.8944}          \\ \bottomrule
\end{tabular}
\end{table}

\textbf{Sea Ice Extent Visualization and Evaluation.} To further quantify and visualize the model's ability at delineating the sea ice edge, which was qualitatively observed in the bias maps, we now evaluate its performance on the Sea Ice Extent (SIE) task. Fig. \ref{fig:SIE1} compares the SIE of FH-Mamba, VMRNN, and FCNet, presenting the results for September 1st, 5th, 10th, and 14th from 2016 to 2020. SIE is defined as the area where the sea ice concentration is at least 15\%, a standard threshold for distinguishing between open water and ice-covered regions. To quantitatively assess the model’s ability to accurately classify these conditions, especially in the marginal ice zone, we introduce Intersection over Union (IoU) metrics. IoU specifically measures the structural alignment of the predicted ice extent against the ground truth. As detailed in Table \ref{tab:IoU_Results}, quantitative analysis for each test year from 2016 to 2020 reveals that FH-Mamba consistently achieves the highest IoU, with an average score of 0.8944, outperforming FCNet (0.8855) and VMRNN (0.8359). These metrics clarify the performance gap that is visually obscured by overlapping lines, demonstrating the performance of FH-Mamba in edge preservation and structural consistency.

\textbf{Analysis of Extreme Low-Ice Events.} To further evaluate FH-Mamba's performance, we focused on three extreme climate events with the smallest sea ice extent values: September 2007, September 2012, and September 2020. The model was trained based on the following datasets: for 2007, SIC data from 1988-2006 was used, while for 2012, data from 1988-2011 was utilized. As for the 2020 analysis, we employed the model already described in the previous section. According to observational records, the lowest point of Arctic sea ice extent (3.387 million square kilometers) occurred on September 17, 2012, followed by records on September 16, 2020 (3.818 million square kilometers) and September 18, 2007 (4.14 million square kilometers). Fig. \ref{fig:Sep9} displays the Arctic sea ice extent for specific September dates in three extreme years. The figure compares the prediction results of FH-Mamba, VMRNN and FCNet with the ground truth data. As indicated by the black rectangular box, FH-Mamba demonstrates higher prediction accuracy at the sea ice edge compared to other SOTA models, with its predictions being closely aligned with the ground truth (black line). This is attributed to FH-Mamba's model design: it ensures the effectiveness of sequence and frequency feature fusion via the Hybrid Shuffle Attention (HSA) and achieves temporally consistent predictions by modeling the temporal dependencies of the input sequence. Consequently, this model can more effectively learn the underlying spatiotemporal correlations in the data.

\begin{table}[h]
\caption{Parameter, computational complexity, inference time, and training time comparison of different methods. The \textbf{bold} and \underline{underline} denote the best and second results.}
\centering
\scalebox{0.85}{
\begin{tabular}{c|cccc}
\toprule
\textbf{Method}   & \textbf{Params (M)} & \textbf{FLOPs (G)} & \textbf{Time (ms)} & \textbf{Training (s)} \\ \midrule
ConvLSTM & 32.76 & 763.13 & 59.343 & 42,551.63 \\
PredRNN & 67.62 & 788.01 & 53.888 & 82,912.45 \\
PredRNNv2 & 67.56 & 800.12 & 61.182 & 81,643.89 \\ 
IceNet & 34.54 & 291.05 & 48.500 & 26,450.33 \\
SimVP & \textbf{14.83} & 292.90 & 45.518 & \textbf{6,826.91} \\
SwinLSTM & 48.23 & 331.68 & 53.598 & 38,120.56 \\
VMRNN & 24.98 & \underline{240.43} & 46.491 & 35,210.18 \\
FCNet & 56.66 & 532.63 & 57.200 & \underline{11,959.27} \\
IceMamba & 44.15 &  300.66  & \textbf{38.465} & 23,501.15 \\
\textbf{FH-Mamba (ours)} & \underline{24.87} & \textbf{236.92} & \underline{45.099} & 17,797.63 \\
\bottomrule
\end{tabular}}
\label{tab:efficiency}
\end{table}

\subsection{Model Complexity and Inference Efficiency Comparison}

Table \ref{tab:efficiency} reports the comparison of model parameters, computational cost (FLOPs), training time, and inference time among different prediction methods. The proposed FH-Mamba achieves a trade-off between model size and efficiency. It contains 24.87M parameters and 236.92G FLOPs, which is the lowest computational cost among all methods. Moreover, FH-Mamba records the competitive inference speed of 45.099 ms, outperforming both convolutional–recurrent models (ConvLSTM, PredRNN, PredRNNv2, and IceNet) and Transformer-based methods (SwinLSTM). In terms of training overhead, FH-Mamba requires 17,797.63 s. It is worth noting that due to the early stopping mechanism, the total training time is influenced by convergence rates. Some models require more epochs to converge, resulting in longer training duration despite their single-step efficiency.

Although SimVP shows the smallest parameter count (14.83 M), its prediction accuracy is notably lower (as shown in subsequent tables). FH-Mamba, in contrast, achieves both high accuracy and high efficiency, demonstrating its effectiveness in capturing spatiotemporal dynamics while maintaining lightweight and real-time inference performance for Arctic sea-ice forecasting.

\subsection{Ablation Studies}

We first conducted comprehensive ablation studies to evaluate the contribution of each component in our FH-Mamba framework. Then, we analyze different scanning directions in 3D Hilbert  and the impact of the HSA module.

\textbf{Effectiveness of Model Components.} The ablation study results shown in Table \ref{tab:abl-1} evaluate the effectiveness of three key components in the FH-Mamba framework: the 3D Hilbert Scanning Strategy (3DHSS), Wavelet transform, and the Hybrid Shuffle Attention (HSA) module. Using only 3DHSS yields better performance than using Wavelet or HSA alone, indicating that the 3D Hilbert scan mechanism is an integral component of our model. The model with all three components achieves the best overall performance with the lowest RMSE and MAE values. It confirms that each module contributes to performance gains and that their combination is complementary.

\begin{table}[ht!]
\centering
\caption{Ablation studies on different modules.}
\begin{tabular}{ccc|cc}
\toprule
\textbf{3DHSS} & \textbf{Wavelet} & \textbf{HSA} & \textbf{RMSE (\%)}$\downarrow$ & \textbf{MAE (\%)}$\downarrow$ \\ \midrule
 &  &  & 8.109 & 2.831 \\
\checkmark &  &  & 6.748 & 2.117 \\
 & \checkmark &  & 6.795 & 2.279 \\
 &  & \checkmark & 6.825 & 2.122 \\
\checkmark & \checkmark & \checkmark & \textbf{6.683} & \textbf{2.061}\\
\bottomrule
\end{tabular}
\label{tab:abl-1}
\end{table}

While the numerical improvements from individual modules on the global RMSE metric may seem modest, this reflects a core challenge of the SIC prediction task. The vast majority of the Arctic region consists of stable thick ice or open water, where most models can achieve high accuracy. Consequently, global metrics are already high and tend to saturate. The true difficulty lies in the highly dynamic and complex marginal ice zones, which are the primary source of prediction errors. Our key components, such as the wavelet branch for high-frequency boundaries and 3DHSS for spatiotemporal locality, are specifically designed to address these challenging regions. Although these zones constitute a smaller portion of the total area, accurately forecasting their evolution is critical for practical applications like climate science and safe navigation. As visualized in Fig. \ref{fig:SIE1}, these modest numerical gains translate into significantly improved preservation of fine spatial structures and boundary details in these critical areas.

\textbf{Impact of 3D Scanning Directions.} Effective scanning is a key issue for Mamba when handling three-dimensional data. We compare five scanning directions on prediction performance in Table \ref{tab:abl-2}. We use the global scanning as in vanilla Mamba, Z-order scanning, Peano curve scanning, spatial-first Hilbert scanning, and temporal-first Hilbert scanning. The global scan in vanilla Mamba results in the worst performance, with an RMSE of 7.041\%, indicating its limited ability to preserve local spatio-temporal structures. In contrast, the Z-order scan achieves better results because it comprehensively considers both spatial and temporal correlations, even though it still suffers from dimensional discontinuities (fractures). The Peano curve further outperforms Z-order as it not only maintains spatio-temporal correlations but also eliminates these dimensional fractures. Ultimately, the Hilbert-based scanning routes achieve the best performance because they provide even stronger spatio-temporal locality than the Peano curve. Moreover, prioritizing temporal locality (temporal-first Hilbert) offers a slight advantage, suggesting that modeling temporal continuity first helps capture the dynamic evolution of sea ice more effectively.

\begin{table}[h]
\centering
\caption{Analysis of scanning directions. (Including estimated Z-order and Peano)}
\begin{tabular}{c|ccc}
\toprule
\textbf{Method}   & \textbf{RMSE (\%)} $\downarrow$ & \textbf{MAE} (\%)$\downarrow$  \\ \midrule
Vanilla global scan & 7.041 & 2.292  \\
Z-order scan & 6.915 & 2.210 \\
Peano curve scan & 6.755 & 2.107 \\
Spatial-first Hilbert & 6.689 & 2.068 \\
Temporal-first Hilbert & \textbf{6.683} & \textbf{2.061} \\ \bottomrule
\end{tabular}
\label{tab:abl-2}
\end{table}

\textbf{Effectiveness of the Feature Fusion Strategy.} Table \ref{tab:abl-3} compares the performance of different fusion strategies for 3D Hilbert sequences and frequency features, including element-wise summation (termed Sum), channel-wise attention gating (termed CAGate), and our HSA. CAGate and HSA perform better than Sum, indicating that attention-based models better capture the interactions between sequence and frequency features. Our HSA achieves the best performance in terms of both RMSE and MAE. It effectively captures the complementary information between sequence and frequency features, leading to more accurate Arctic SIC predictions.

\begin{table}[h]
\centering
\caption{Performance comparison of feature fusion.}
\begin{tabular}{c|cc}
\toprule
\textbf{Method} & \textbf{RMSE (\%)}$\downarrow$ & \textbf{MAE (\%)}$\downarrow$ \\ \midrule
Sum & 7.017 & 2.151 \\
CAGate & 6.775 & 2.120 \\
HSA (ours) & \textbf{6.683} & \textbf{2.061} \\ \bottomrule
\end{tabular}
\label{tab:abl-3}
\end{table}

\textbf{Justification for the Wavelet Transform Branch.}
A key challenge in SIC prediction is capturing the fine-grained details of marginal ice zones, which correspond to high-frequency information in the spatial domain. To validate that our choice of a Discrete Wavelet Transform (DWT) branch is effective for this purpose, we conducted a comparative analysis against several alternative designs: 1) a baseline model without any specialized frequency branch, 2) a branch using simple Multi-Scale Convolutions (MSC) with different kernel sizes to capture features at various scales, and 3) a STFT-based Spectral Attention (STFT-SA) module that operates in the Fourier domain. As shown in Table \ref{tab:abl-4}, while both MSC and STFT-SA offer improvements over the baseline, our DWT-based approach achieves the lowest error. This suggests that the multi-resolution analysis capability of DWT is particularly effective at decomposing and emphasizing the precise boundary details critical for SIC forecasting, justifying its selection over other frequency and multi-scale analysis techniques.

\begin{table}[h]
\centering
\caption{Performance comparison of different high-frequency feature extraction branches. Our DWT-based approach yields the best results.}
\begin{tabular}{c|cc}
\toprule
\textbf{Method} & \textbf{RMSE (\%)}$\downarrow$ & \textbf{MAE (\%)}$\downarrow$ \\ \midrule
None (Baseline)      & 6.713              & 2.092             \\
MSC (Multi-scale Conv) & 6.701              & 2.084             \\
STFT-SA              & 6.692              & 2.072             \\
DWT (Ours)  & \textbf{6.683}     & \textbf{2.061} \\ \bottomrule
\end{tabular}
\label{tab:abl-4}
\end{table}

\textbf{Impact of Different Wavelet Bases.} To justify our choice of the wavelet basis function in the FSSM, we conducted a comparative analysis using three different wavelet families: Haar, Daubechies (db2), and Biorthogonal (bior1.3). The Haar wavelet is known for its capability to capture sudden transitions, making it suitable for detecting sea ice edges. Daubechies and Biorthogonal wavelets offer higher vanishing moments and different symmetry properties.

\begin{table}[h]
\centering
\caption{Performance comparison of different wavelet bases.}
\label{tab:wavelet_bases}
\begin{tabular}{c|cc}
\toprule
\textbf{Wavelet Basis} & \textbf{RMSE (\%)} & \textbf{MAE (\%)} \\
\midrule
Daubechies (db2) & 6.691 & 2.068 \\
Biorthogonal (bior1.3) & 6.689 & 2.065 \\
Haar (Ours) & \textbf{6.683} & \textbf{2.061} \\
\bottomrule
\end{tabular}
\end{table}

As shown in Table \ref{tab:wavelet_bases}, the performance differences among these bases are negligible. The RMSE fluctuates within a narrow range of 0.01\%, and MAE within 0.007\%. This suggests that the proposed FH-Mamba is robust to the selection of wavelet bases. We adopted the Haar wavelet in our final model primarily due to its computational simplicity and effectiveness in preserving high-frequency boundary information without introducing additional computational overhead.

\subsection{Parameter Sensitivity Analysis}

\begin{figure}[t!]
    \centering
    \includegraphics[width=0.8\linewidth]{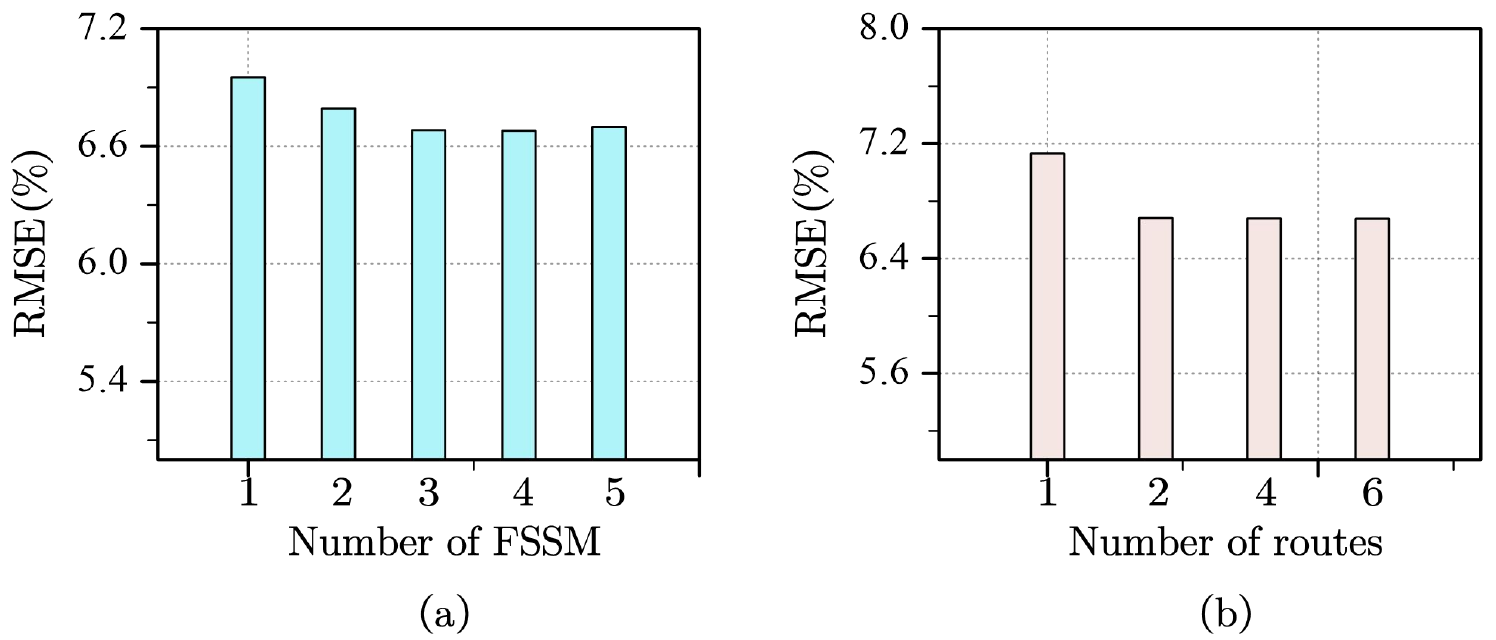}
    \caption{Parameter analysis. (a) Relationship between the number of FSSM modules and the corresponding RMSE. (b) Relationship between the number of routes in 3D Hilbert scanning and the corresponding RMSE.}
    \label{fig:para}
\end{figure}

\textbf{Impact of the Number of FSSM Blocks.} The number of Frequency-enhanced State Space Modules (FSSMs) is an important parameter that may affect the performance of Arctic Sea Ice Concentration (SIC) prediction. In this study, we have conducted extensive experiments to study FSSM, and the results are shown in Fig. \ref{fig:para}(a). As the number of FSSM modules increases from 1 to 5, the RMSE consistently decreases. This trend indicates that incorporating more FSSM modules improves prediction accuracy. However, we observe a clear law of diminishing returns beyond three modules. The incremental accuracy gain from 3 to 5 modules is negligible compared to the increase in computational complexity. In light of this efficiency-accuracy trade-off, we deliberately select 3 modules as the optimal configuration. This choice ensures the model retains its prediction capabilities while remaining sufficiently lightweight for the rapid processing requirements of large-scale Arctic spatiotemporal data.

\textbf{Impact of the Number of 3D Hilbert Scan Routes.} The 3D Hilbert scan can have multiple valid paths, depending on the configuration and transformation rules used during construction. In practice, after generating the 3D Hilbert curve, we can obtain two scanning paths: a forward scan and a backward scan. Next, by rotating the input 3D data, we can obtain more scanning routes. In Fig. \ref{fig:para}(b), the bar chart shows the effect of varying the number of 3D Hilbert scanning routes on RMSE for prediction performance. When we use two routes, the RMSE significantly decreases compared with using only 1 route. However, when the number of routes increases beyond 2, the RMSE remains nearly unchanged. Therefore, in our implementations, we use 2 routes for 3D Hilbert scanning.

\textbf{Long-Term Iterative Forecasting.} To predict longer SIC sequences, we explore the performance of FH-Mamba for recursive prediction. Specifically, the predicted results for the first 14 days are used as the inputs of FH-Mamba to predict the SIC for the next 14 days. Likewise, the predicted SIC for the next 14 days can be used as inputs to generate a 42-day SIC prediction. The experimental results for recursive prediction are shown in Table \ref{tab:recursive}. It can be observed that when the number of recursive steps increases (extending the prediction horizon to 28, 42, and 56 days), the prediction performance gradually degrades. This performance decline is expected due to the accumulation of prediction errors as the model repeatedly uses its own outputs as inputs for subsequent predictions. Despite this, for 56-day SIC prediction, the MAE increases to 3.697\%, revealing limitations for precise navigation due to recursive error accumulation. While the model captures the general evolution, its reliability for real-world applications decreases at longer horizons.

\begin{table}[h]
\caption{Performance of our FH-Mamba with different recursive steps.}
\centering
\begin{tabular}{cc|ccc}
\toprule
\makecell{\textbf{Recursive}\\\textbf{step}} & \makecell{\textbf{Prediction}\\\textbf{period}} & \textbf{RMSE}$\downarrow$ & \textbf{MAE}$\downarrow$ & \textbf{NSE}$\uparrow$ \\
\midrule
0 & 14 days & \textbf{6.683} & \textbf{2.061} & \textbf{96.609} \\
1 & 28 days & 9.616  & 3.064 & 87.569 \\
2 & 42 days & 11.462 & 3.690 & 81.177 \\
3 & 56 days & 11.634 & 3.697 & 81.166 \\ 
\bottomrule
\end{tabular}
\label{tab:recursive}
\end{table}

\subsection{Uncertainty Quantification}

While our proposed FH-Mamba is designed as a deterministic model providing point forecasts, real-world applications such as ensuring safe navigation often benefit from understanding the model's prediction uncertainty. To this end, we extend our framework to enable probabilistic forecasting, allowing it to quantify the uncertainty associated with its predictions.

This extension is achieved by modifying the model's output layer. Instead of predicting a single value for each grid point, the probabilistic version is trained to predict the parameters of a Gaussian distribution: the mean \( \mu \) and the standard deviation \( \sigma \). The predicted mean \( \mu \) represents the most likely SIC value, analogous to the output of our deterministic model. The standard deviation \( \sigma \) represents the model's uncertainty; a larger \( \sigma \) indicates lower confidence in the prediction for that specific point. To ensure that \( \sigma \) is always positive, the network's raw output for the standard deviation is passed through a softplus activation function. For training this probabilistic model, we replace the L1 loss with the Negative Log-Likelihood (NLL) loss, which is a standard objective for learning probability distributions. The goal is to maximize the likelihood of the ground truth data given the predicted distribution.

The experimental results are presented in Table \ref{tab:uncertainty_results}. As can be seen, the probabilistic version of FH-Mamba experiences a negligible increase in RMSE and MAE compared to its deterministic counterpart. This demonstrates that the model can provide valuable uncertainty information without significantly compromising its point prediction accuracy, enhancing its practical utility for critical decision-making processes.

\begin{table}[h!]
\centering
\caption{Performance comparison between the deterministic and probabilistic versions of FH-Mamba.}
\label{tab:uncertainty_results}
\begin{tabular}{lccc}
\toprule
\textbf{Model Version} & \textbf{RMSE (\%)$\downarrow$} & \textbf{MAE (\%)$\downarrow$} & \textbf{NLL$\downarrow$} \\
\midrule
Deterministic & 6.683 & 2.061 & N/A \\
Probabilistic & 6.712 & 2.075 & -2.453 \\
\bottomrule
\end{tabular}
\end{table}

\section{Conclusion}

In this paper, we propose FH-Mamba, a novel framework for short-term Arctic SIC prediction. We focus on two challenges:

For the first challenge: \textbf{\textit{How can we effectively model temporal correlations by using SSMs?}} To address this, we employ 3D Hilbert scanning mechanism that enhances the model’s ability to learn spatiotemporal patterns. Unlike conventional global scans, the scanning in our FH-Mamba is performed across both temporal and spatial dimensions simultaneously. It enables the model to better capture local temporal correlations and subtle transitions in sea ice evolution. Experimental comparisons with SOTA methods demonstrate that our FH-Mamba effectively improves the Arctic SIC prediction. The model's performance improvements were consistently demonstrated on both the OSI-450a1 and AMSR2 datasets.

For the second challenge: \textbf{\textit{How can we enhance the details of the Arctic margin regions?}} To tackle this, we employ wavelet transform to amplify the high-frequency components. Additionally, we design a HSA module to further exploit the complementary information between sequence features and frequency features. HSA performs channel-wise attention to adaptively fuse these features, allowing the model to better capture complex dependencies across both domains. Visualization results of predicted SIC maps clearly show that FH-Mamba produces sharper and more accurate boundary details, particularly in marginal ice zones.

Overall, FH-Mamba provides a method for fine-grained and high-fidelity Arctic SIC forecasting, offering valuable insights for future developments in physics-aware and frequency-enhanced SSM-based ocean prediction models.

\bibliography{source}

@article{parkinson2021sea,
  title={Sea ice extents continue to set new records: Arctic, Antarctic, and global results},
  author={Parkinson, Claire L and DiGirolamo, Nicolo E},
  journal={Remote Sensing of Environment},
  volume={267},
  pages={112753},
  year={2021}}

@ARTICLE{survey2022,
  author={Wang, Senzhang and Cao, Jiannong and Yu, Philip S.},
  journal={IEEE Transactions on Knowledge and Data Engineering}, 
  title={Deep Learning for Spatio-Temporal Data Mining: A Survey}, 
  year={2022},
  volume={34},
  number={8},
  pages={3681-3700}}

@article{survey2025,
title = {Spatio-temporal prediction using graph neural networks: A survey},
journal = {Neurocomputing},
volume = {643},
pages = {130400},
year = {2025},
author = {Vincenzo Capone and Angelo Casolaro and Francesco Camastra},
pages = {1-18}}

@article{hilbert1891ueber,
  title={Ueber die stetige Abbildung einer Linie auf ein Fl{\"a}chenst{\"u}ck},
  author={Hilbert, David},
  journal={Mathematische Annalen},
  volume={38},
  number={3},
  pages={459--460},
  year={1891},
  publisher={Springer}
}

@INPROCEEDINGS{swintransformer2021,
  author={Liu, Ze and Lin, Yutong and Cao, Yue and Hu, Han and Wei, Yixuan and Zhang, Zheng and Lin, Stephen and Guo, Baining},
  booktitle={2021 IEEE/CVF International Conference on Computer Vision (ICCV)}, 
  title={Swin Transformer: Hierarchical Vision Transformer using Shifted Windows}, 
  year={2021},
  volume={},
  number={},
  pages={9992-10002},
  doi={10.1109/ICCV48922.2021.00986}}

@InProceedings{unet2015CoRR,
author={Olaf Ronneberger and Philipp Fischer and Thomas Brox},
title={{U-Net}: Convolutional Networks for Biomedical Image Segmentation},
booktitle={Proceedings of Medical Image Computing and Computer-Assisted Intervention (MICCAI)},
year={2015},
pages={234--241}}

@inproceedings{NIPS2017_transformer,
    author = {Vaswani, Ashish and Shazeer, Noam and Parmar, Niki and Uszkoreit, Jakob and Jones, Llion and Gomez, Aidan N and Kaiser, \L ukasz and Polosukhin, Illia},
     booktitle = {Proceedings of Advances in Neural Information Processing Systems (NeurIPS)},
     pages = {1-13},
     title = {Attention is All you Need},
     volume = {30},
     year = {2017}}

@article{Spreen2008SeaIR,
  title={Sea ice remote sensing using AMSR‐E 89‐GHz channels},
  author={Gunnar Spreen and Lars Kaleschke and Georg C. Heygster},
  journal={Journal of Geophysical Research},
  year={2008},
  volume={113}}

@Article{icenet21,
author={Andersson, Tom R. and Hosking, J. Scott and P{\'e}rez-Ortiz, Mar{\'i}a and Paige, Brooks and Elliott, Andrew and Russell, Chris and Law, Stephen and Jones, Daniel C. and Wilkinson, Jeremy and Phillips, Tony and Byrne, James and Tietsche, Steffen and Sarojini, Beena Balan and Blanchard-Wrigglesworth, Eduardo and Aksenov, Yevgeny and Downie, Rod and Shuckburgh, Emily},
title={Seasonal Arctic sea ice forecasting with probabilistic deep learning},
journal={Nature Communications},
year={2021},
volume={12},
number={1},
pages={1--12}}

@article{chi2017prediction,
  title={Prediction of Arctic sea ice concentration using a fully data driven deep neural network},
  author={Chi, Junhwa and Kim, Hyun-Choel},
  journal={Remote Sensing},
  volume={9},
  number={12},
  pages={1-22},
  year={2017}}

@article{dee2011era,
  title={The ERA-Interim reanalysis: Configuration and performance of the data assimilation system},
  author={Dee, Dick P and Uppala, S Mꎬ and Simmons, Adrian J and Berrisford, Paul and Poli, Paul and Kobayashi, Shinya and Andrae, U and Balmaseda, MA and Balsamo, G and Bauer, d P and others},
  journal={Journal of the Royal Meteorological Society},
  volume={137},
  number={656},
  pages={553--597},
  year={2011}}

@inproceedings{gao2022simvp,
  title={SimVP: Simpler yet better video prediction},
  author={Gao, Zhangyang and Tan, Cheng and Wu, Lirong and Li, Stan Z},
  booktitle={IEEE/CVF Conference on Computer Vision and Pattern Recognition (CVPR)},
  pages={3170--3180},
  year={2022}}

@InProceedings{mamba24,
  title={Mamba: Linear-time sequence modeling with selective state spaces},
  author={Gu, Albert and Dao, Tri},
  booktitle = {Proceedings of the Conference on Language Modeling (COLM)},
  year={2024},
  pages = {1-23}}

@article{guemas2016review,
  title={A review on Arctic sea-ice predictability and prediction on seasonal to decadal time-scales},
  author={Guemas, Virginie and Blanchard-Wrigglesworth, Edward and Chevallier, Matthieu and Day, Jonathan J and D{\'e}qu{\'e}, Michel and Doblas-Reyes, Francisco J and Fu{\v{c}}kar, Neven S and Germe, Agathe and Hawkins, Ed and Keeley, Sarah and others},
  journal={Quarterly Journal of the Royal Meteorological Society},
  volume={142},
  number={695},
  pages={546--561},
  year={2016}}

@InProceedings{mambavision2025CVPR,
    author    = {Hatamizadeh, Ali and Kautz, Jan},
    title     = {MambaVision: A hybrid Mamba-Transformer vision backbone},
    booktitle = {Proceedings of the Computer Vision and Pattern Recognition Conference (CVPR)},
    month     = {June},
    year      = {2025},
    pages     = {25261-25270}}

@article{horvath2020bayesian,
  title={A Bayesian logistic regression for probabilistic forecasts of the minimum September Arctic sea ice cover},
  author={Horvath, Sean and Stroeve, Julienne and Rajagopalan, Balaji and Kleiber, William},
  journal={Earth and Space Science},
  volume={7},
  number={10},
  pages={1-18},
  year={2020}}

@article{hung2010atmospheric,
  title={Atmospheric monitoring of organic pollutants in the Arctic under the Arctic Monitoring and Assessment Programme (AMAP): 1993--2006},
  author={Hung, Hayley and Kallenborn, Roland and Breivik, Knut and Su, Yushan and Brorstr{\"o}m-Lund{\'e}n, Eva and Olafsdottir, Kristin and Thorlacius, Johanna M and Lepp{\"a}nen, Sirkka and Bossi, Rossana and Skov, Henrik and others},
  journal={Science of the Total Environment},
  volume={408},
  number={15},
  pages={2854--2873},
  year={2010}}

@article{kim2020prediction,
  title={Prediction of monthly Arctic sea ice concentrations using satellite and reanalysis data based on convolutional neural networks},
  author={Kim, Young Jun and Kim, Hyun-Cheol and Han, Daehyeon and Lee, Sanggyun and Im, Jungho},
  journal={The Cryosphere},
  volume={14},
  number={3},
  pages={1083--1104},
  year={2020}}

@article{lavergne2019version,
  title={Version 2 of the EUMETSAT OSI SAF and ESA CCI sea-ice concentration climate data records},
  author={Lavergne, Thomas and S{\o}rensen, Atle Macdonald and Kern, Stefan and Tonboe, Rasmus and Notz, Dirk and Aaboe, Signe and Bell, Louisa and Dybkj{\ae}r, Gorm and Eastwood, Steinar and Gabarro, Carolina and others},
  journal={The Cryosphere},
  volume={13},
  number={1},
  pages={49--78},
  year={2019}}

@InProceedings{efficientVIM2025CVPR,
    author    = {Lee, Sanghyeok and Choi, Joonmyung and Kim, Hyunwoo J.},
    title     = {EfficientViM: Efficient vision Mamba with hidden state mixer based state space duality},
    booktitle = {Proceedings of the Computer Vision and Pattern Recognition Conference (CVPR)},
    month     = {June},
    year      = {2025},
    pages     = {14923-14933}}

@InProceedings{li25cvpr,
    author    = {Li, Boyun and Zhao, Haiyu and Wang, Wenxin and Hu, Peng and Gou, Yuanbiao and Peng, Xi},
    title     = {MaIR: A locality- and continuity-preserving Mamba for image restoration},
    booktitle = {Proceedings of the Computer Vision and Pattern Recognition Conference (CVPR)},
    month     = {June},
    year      = {2025},
    pages     = {7491-7501}}

@InProceedings{defmamba2025CVPR,
    author    = {Liu, Leiye and Zhang, Miao and Yin, Jihao and Liu, Tingwei and Ji, Wei and Piao, Yongri and Lu, Huchuan},
    title     = {DefMamba: Deformable visual state space model},
    booktitle = {Proceedings of the Computer Vision and Pattern Recognition Conference (CVPR)},
    month     = {June},
    year      = {2025},
    pages     = {8838-8847}}

@InProceedings{GroupMamba2025CVPR,
    author    = {Shaker, Abdelrahman and Wasim, Syed Talal and Khan, Salman and Gall, Juergen and Khan, Fahad Shahbaz},
    title     = {GroupMamba: Efficient group-based visual state space model},
    booktitle = {Proceedings of the Computer Vision and Pattern Recognition Conference (CVPR)},
    month     = {June},
    year      = {2025},
    pages     = {14912-14922}}

@InProceedings{icediff25,
    author    = {Xu, Jingyi and Tu, Siwei and Yang, Weidong and Fei, Ben and Li, Shuhao and Liu, Keyi and Luo, Yeqi and Ma, Lipeng and Bai, Lei},
    title     = {IceDiff: High resolution and high-quality Arctic sea ice Forecasting with generative diffusion prior},
    booktitle = {Proceedings of the Computer Vision and Pattern Recognition Conference (CVPR)},
    month     = {June},
    year      = {2025},
    pages     = {10567-10576}}

@article{fcnet25,
  author={Zhang, Jialiang and Gao, Feng and Gan, Yanhai and Dong, Junyu and Du, Qian},
  journal={IEEE Transactions on Geoscience and Remote Sensing}, 
  title={Frequency-compensated network for daily Arctic sea ice concentration prediction}, 
  year={2025},
  volume={63},
  pages={1-15}}

@inproceedings{mamba24cikm,
author = {Zhang, Hanqi and Chen, Chong and Mei, Lang and Liu, Qi and Mao, Jiaxin},
title = {Mamba Retriever: Utilizing Mamba for effective and efficient dense retrieval},
year = {2024},
booktitle = {Proceedings of the 33rd ACM International Conference on Information and Knowledge Management (CIKM)},
pages = {4268–4272}}

@InProceedings{zhou25cvpr,
    author    = {Zhou, Shiyang and Zeng, Haijin and Lu, Yunfan and Shao, Tong and Tang, Ke and Chen, Yongyong and Liu, Jie and Su, Jingyong},
    title     = {Binarized Mamba-Transformer for lightweight quad bayer HybridEVS demosaicing},
    booktitle = {Proceedings of the Computer Vision and Pattern Recognition Conference (CVPR)},
    month     = {June},
    year      = {2025},
    pages     = {8817-8827}}

@article{yang2019improving,
  title={Improving Arctic sea ice seasonal outlook by ensemble prediction using an ice-ocean model},
  author={Yang, Qinghua and Mu, Longjiang and Wu, Xingren and Liu, Jiping and Zheng, Fei and Zhang, Jinlun and Li, Chuanjin},
  journal={Atmospheric research},
  volume={227},
  pages={14--23},
  year={2019},
  publisher={Elsevier}
}

@article{tepes2021changes,
  title={Changes in elevation and mass of Arctic glaciers and ice caps, 2010--2017},
  author={Tepes, Paul and Gourmelen, Noel and Nienow, Peter and Tsamados, M and Shepherd, A and Weissgerber, Flora},
  journal={Remote Sensing of Environment},
  volume={261},
  pages={112481},
  year={2021},
  publisher={Elsevier}
}

@article{ren2022data,
  title={A data-driven deep learning model for weekly sea ice concentration prediction of the Pan-Arctic during the melting season},
  author={Ren, Yibin and Li, Xiaofeng and Zhang, Wenhao},
  journal={IEEE Transactions on Geoscience and Remote Sensing},
  volume={60},
  pages={1--19},
  year={2022},
  publisher={IEEE}
}

@INPROCEEDINGS{tang2024vmrnn,
  author={Tang, Yujin and Dong, Peijie and Tang, Zhenheng and Chu, Xiaowen and Liang, Junwei},
  booktitle={2024 IEEE/CVF Conference on Computer Vision and Pattern Recognition Workshops (CVPRW)}, 
  title={VMRNN: Integrating Vision Mamba and LSTM for Efficient and Accurate Spatiotemporal Forecasting}, 
  year={2024},
  volume={},
  number={},
  pages={5663-5673},
  doi={10.1109/CVPRW63382.2024.00575}}

@INPROCEEDINGS{tang2023swinlstm,
  author={Tang, Song and Li, Chuang and Zhang, Pu and Tang, RongNian},
  booktitle={2023 IEEE/CVF International Conference on Computer Vision (ICCV)}, 
  title={SwinLSTM: Improving Spatiotemporal Prediction Accuracy using Swin Transformer and LSTM}, 
  year={2023},
  volume={},
  number={},
  pages={13424-13433},
  doi={10.1109/ICCV51070.2023.01239}}

@inproceedings{wang17predrnn,
 author = {Wang, Yunbo and Long, Mingsheng and Wang, Jianmin and Gao, Zhifeng and Yu, Philip S},
 booktitle = {Advances in Neural Information Processing Systems (NeurIPS)},
 editor = {I. Guyon and U. Von Luxburg and S. Bengio and H. Wallach and R. Fergus and S. Vishwanathan and R. Garnett},
 pages = {1-13},
 title = {PredRNN: Recurrent neural networks for predictive learning using spatiotemporal LSTMs},
 volume = {30},
 year = {2017}}

@article{wang2022predrnn,
  title={PredRNN: A recurrent neural network for spatiotemporal predictive learning},
  author={Wang, Yunbo and Wu, Haixu and Zhang, Jianjin and Gao, Zhifeng and Wang, Jianmin and Philip, S Yu and Long, Mingsheng},
  journal={IEEE Transactions on Pattern Analysis and Machine Intelligence},
  volume={45},
  number={2},
  pages={2208--2225},
  year={2022}}

@article{stroeve2018changing,
  title={Changing state of Arctic sea ice across all seasons},
  author={Stroeve, Julienne and Notz, Dirk},
  journal={Environmental Research Letters},
  volume={13},
  number={10},
  pages={103001},
  year={2018},
  publisher={IOP Publishing}
}

@article{sicnet25gmd,
AUTHOR = {Ren, Y. and Li, X. and Wang, Y.},
TITLE = {SICNet V1.0: a Transformer-based deep learning model for seasonal Arctic sea ice prediction by incorporating sea ice thickness data},
JOURNAL = {Geoscientific Model Development},
VOLUME = {18},
YEAR = {2025},
NUMBER = {9},
PAGES = {2665--2678}}

@Article{Wang2025,
author={Wang, Wei
and Yang, Weidong
and Wang, Lei
and Wang, Guihua
and Lei, Ruibo},
title={Seasonal forecasting of Pan-Arctic sea ice with state space model},
journal={npj Climate and Atmospheric Science},
year={2025},
month={May},
day={07},
volume={8},
number={1},
pages={172}}

@inproceedings{rainmamba24,
author = {Wu, Hongtao and Yang, Yijun and Xu, Huihui and Wang, Weiming and Zhou, Jinni and Zhu, Lei},
title = {RainMamba: Enhanced locality learning with state space models for video deraining},
year = {2024},
booktitle = {ACM Multimedia},
pages = {7881–7890}}

@inproceedings{vim24icml,
title={Vision Mamba: Efficient visual representation learning with bidirectional state space model},
author={Zhu, Lianghui and Liao, Bencheng and Zhang, Qian and Wang, Xinlong and Liu, Wenyu and Wang, Xinggang},
booktitle={Proceedings of International Conference on Machine Learning (ICML)},
year = {2024},
pages = {1-13}}

@article{chen21esa,
title = {All-nearest-neighbors finding based on the Hilbert curve},
author = {Hue-Ling Chen and Ye-In Chang},
journal = {Expert Systems with Applications},
volume = {38},
number = {6},
pages = {7462-7475},
year = {2011}}

@article{keller22,
      title={Rendering along the Hilbert Curve}, 
      author={Alexander Keller and Carsten Wächter and Nikolaus Binder},
      year={2022},
      journal={arXiv: 2207.05415},
      pages = {1-13}}

@inproceedings{uddin18,
author = {Uddin, Reaz and Ravishankar, Chinya V. and Tsotras, Vassilis J.},
title = {Indexing moving object trajectories with hilbert curves},
year = {2018},
pages = {416–419},
numpages = {4},
location = {Seattle, Washington}}

@inproceedings{liang24pointmamba,
      title={PointMamba: A Simple State Space Model for Point Cloud Analysis}, 
      author={Liang, Dingkang and Zhou, Xin and Xu, Wei and Zhu, Xingkui and Zou, Zhikang and Ye, Xiaoqing and Tan, Xiao and Bai, Xiang},
      booktitle={Advances in Neural Information Processing Systems (NeurIPS)},
      year={2024}}

@article{moon01tkde,
  author={Moon, B. and Jagadish, H.V. and Faloutsos, C. and Saltz, J.H.},
  journal={IEEE Transactions on Knowledge and Data Engineering}, 
  title={Analysis of the clustering properties of the Hilbert space-filling curve}, 
  year={2001},
  volume={13},
  number={1},
  pages={124-141}}

@article{butz71tc,
  author={Butz, A.R.},
  journal={IEEE Transactions on Computers}, 
  title={Alternative Algorithm for Hilbert's Space-Filling Curve}, 
  year={1971},
  volume={C-20},
  number={4},
  pages={424-426}}

@article{cjf22tip,
  author={Chen, Jiafeng and Yu, Lu and Wang, Wenyi},
  journal={IEEE Transactions on Image Processing}, 
  title={Hilbert Space Filling Curve Based Scan-Order for Point Cloud Attribute Compression}, 
  year={2022},
  volume={31},
  number={},
  pages={4609-4621}}

@article{wj18tvcg,
  author={Weissenböck, Johannes and Fröhler, Bernhard and Gröller, Eduard and Kastner, Johann and Heinzl, Christoph},
  journal={IEEE Transactions on Visualization and Computer Graphics}, 
  title={Dynamic Volume Lines: Visual Comparison of 3D Volumes through Space-filling Curves}, 
  year={2019},
  volume={25},
  number={1},
  pages={1040-1049}}

@article{xyc25grsl,
  author={Xu, Yichu and Han, Chengxi and Chen, Shi and Jin, Yao and Miao, Yuchun and Guo, Haonan and Wang, Di},
  journal={IEEE Geoscience and Remote Sensing Letters}, 
  title={PHDMamba: Progressive Hybrid Mamba for Hyperspectral Image Classification}, 
  year={2025},
  volume={22},
  pages={1-5}}

@article{llw25tgrs,
  author={Li, Linwei and Wang, Bin},
  journal={IEEE Transactions on Geoscience and Remote Sensing}, 
  title={DPMN: Deep Prior Mamba Network for Hyperspectral Anomaly Detection}, 
  year={2025},
  volume={63},
  pages={1-16}}

@article{llh25tgrs,
  author={Liang, Lianhui and Zhang, Jing and Duan, Puhong and Kang, Xudong and Xinzhang Wu, Thomas and Li, Jun and Plaza, Antonio},
  journal={IEEE Transactions on Geoscience and Remote Sensing}, 
  title={LKMA: Learnable Kernel and Mamba With Spatial–Spectral Attention Fusion for Hyperspectral Image Classification}, 
  year={2025},
  volume={63},
  pages={1-14}}

@article{zzh25jstars,
  author={Zhang, Zuoheng and Hu, Zhengsheng and Cao, Binghan and Li, Pei and Su, Qun and Dong, Zhao and Wang, Tao},
  journal={IEEE Journal of Selected Topics in Applied Earth Observations and Remote Sensing}, 
  title={Wiener Filter-Based Mamba for Remote Sensing Image Super-Resolution With Novel Degradation}, 
  year={2025},
  volume={18},
  pages={26295-26308}}
\bibliographystyle{IEEEtran}

\end{document}